\documentstyle[12pt,epsf,epsfig,a4]{article}
\begin{document}
\begin {titlepage}
\begin{flushleft}
FSUJ TPI QO-6/98
\end{flushleft}
\begin{flushright}
March, 1998
\end{flushright}
\vspace{20mm}
\begin{center}
{\Large {\bf
Quantum state engineering using conditional measurement on a beam
splitter}\\[3ex]
\Large M. Dakna, L. Kn\"oll, D.--G.  Welsch}\\[3ex]
Friedrich-Schiller-Universit\"{a}t Jena,
Theoretisch-Physikalisches Institut
\\
Max-Wien-Platz 1, D-07743 Jena, Germany\\[2.ex] 
\vspace{25mm}
\end{center}
\begin{center}
\bf{Abstract}
\end{center}
State preparation via conditional output measurement on a beam
splitter is studied, assuming the signal mode is mixed
with a mode prepared in a Fock state and photon numbers are 
measured in one of the output channels. It is shown that
the mode in the other output channel is prepared in either 
a photon-subtracted or a photon-added Jacobi polynomial state, 
depending upon the difference between the number of photons in the 
input Fock state and the number of photons in the output Fock state 
onto which it is projected. 
The properties of the conditional output states are studied
for coherent and squeezed input states, and the probabilities
of generating the states are calculated. Relations to other states,
such as near-photon-number states and squeezed-state-excitations,
are given and proposals are made for generating them by combining
the scheme with others. Finally, effects of realistic photocounting 
and Fock-state preparation are discussed.
\end{titlepage}
\section{Introduction}
\label{sec1}
Over the last years numerous workers have studied various
nonclassical states of radiation and proposed schemes for
producing them. Particular interest has been devoted, e.g., to Fock states 
(for a review, see \cite{Davidovich}) and states derived
from Fock states by coherently displacing and/or squeezing them 
\cite{Boiteux,Roy,Wuensche0,Satya,Nieto1}, 
superpositions of mesoscopically distinguishable
states, such as Schr\"odinger-cat-like states   
(for a review, see \cite{Buzek}),
near-photon-number states (also called crescent states)
\cite{Yuen2,Hradil,Luks1,Luks2,Marian1,Agarwal1}, binomial states 
\cite{Stoler,Lee,Barranco}, inverse binomial states \cite{Barnett,Pegg},
squeezed-state excitations \cite{Wuensche1,Dodonov2} and 
the SU(2) and SU(1,1) minimum-uncertainty states \cite{Luks1,Luis1,Brif1}.
Another interesting class of nonclassical states that have been  
a subject of increasing interest are photon-added and 
photon-subtracted states that are obtained by repeated application of 
photon creation or destruction operators, respectively, on a given 
state \cite{Dakna1,Dakna2,Dakna3,Ban1,Jones1,Agarwal2,Zhang1,Dodonov1}.
Similarly, states obtained by the repeated application 
of the inverse boson operators have also been considered \cite{Mehta1}. 

Despite the large body of work, only few of the above mentioned 
nonclassical states have been generated experimentally so far.
Designing of realistic schemes for generating specific 
quantum states and realization of the schemes in the laboratory 
have been one of the most exciting challenges to the researchers. A promising 
method of quantum state engineering has been conditional measurement,
e.g., generation of a desired state by state reduction in one of two 
entangled quantum objects owing to an appropriate measurement on the 
other object. Typical examples that have been considered for nonclassical state generation
via conditional measurements are the  interfering fields in the output
channels of a beam splitter \cite{Dakna1,Dakna2,Ban1,Ban2,Ban3},
waves produced by parametric amplifiers 
\cite{Luks2,Agarwal1,Luis1,Ban2,Ban4,Song,Yurke3,Watanabe1,Yamamoto1,Ribeiro,Agarwal3} and 
degenerate four-wave mixers \cite{Agarwal1,Ban5} and systems of 
the Jaynes-Cummings type in cavity QED \cite{Brune1,Brune2,VogelK,Parkins1} 
or trapped-ion studies \cite{Cirac1,Monroe1,Monroe2}. Further, state 
reduction via continuous measurement has also been considered 
\cite{Ueda1,Ueda2,Ueda3,Ueda4,Ueda5,Ban6}.

In this paper we study the class of states generated by conditional 
output measurement on a beam splitter in the case when an input mode prepared 
in some quantum state and another input mode prepared in an
$n$ photon Fock state are mixed and in one of the output channels of 
the beam splitter a photon-number measurement yields $m$ photons. We 
show that the conditional output states are 
photon-subtracted ($n$ $\!<$ $\!m$) or photon-added ($n$ $\!>$ $\!m$)
Jacobi polynomial states, i.e., states that are obtained by 
($|n$ $\!-$ $\!m|$ times) repeated 
application of either the photon destruction operator 
or the photon creation operator, respectively, to Jacobi polynomial states.
It is worth noting that the scheme can be used to generate
photon-subtracted and photon-added Jacobi polynomial states for
various classes of input quantum states, such as thermal states, coherent 
states, squeezed states and displaced photon-number states. 
In particular, for $n$ $\!=$ $\!0$ and $m$ $\!=$ $\!0$, respectively, 
ordinary photon-subtracted and photon-added states 
\cite{Dakna1,Dakna2,Dakna3,Ban1,Jones1,Agarwal2,Zhang1,Dodonov1}
are observed.
 
In order to illustrate the nonclassical properties of 
pho\-ton-subtracted and photon-added Jacobi polynomial sta\-tes,
we study them for coherent input states in 
more detail. We analyze the states
in terms of the photon-number and quadrature-component distributions
and the Wigner and Husimi functions, and we calculate
the probability of producing them.  
Further, we briefly address the case of squeezed
vacuum input. It is worth noting that in this case the produced
photon-subtracted and photon-added Jacobi polynomial states
 -- similarly to ordinary photon-subtracted and photon-added
squeezed vacuum states \cite{Dakna1,Dakna3} -- are examples of 
classes of Schr\"odinger-cat-like states.
 
We further study the relation of the conditional output  
states to other classes of nonclassical states. In particular 
we show that near-photon-number states
\cite{Yuen2,Hradil,Luks1,Luks2,Marian1,Agarwal1}
and squeezed-state excitations \cite{Wuensche1,Dodonov2}
can be generated by photon adding and subsequent coherent
displacement and/or squeezing. It also turns out  
that photon-subtracted and photon-added Jacobi
polynomial coherent states are finite superpositions of
ordinary photon-added coherent states, which for themselves  
are finite superpositions of displaced Fock states \cite{Agarwal2}. 
Similarly, photon-subtracted and photon-added squeezed vacuum states
can be regarded as two different finite superpositions of squee\-zed 
number states.

This paper is organized as follows. Section \ref{sec2} presents 
the basic scheme for generation of photon-subtracted and photon-added
Jacobi polynomial states. The properties of the states are studied 
in Sections \ref{sec2a} -- \ref{subsec5b}, with special
emphasis on coherent input states (Section \ref{subsec5a}) and 
squeezed vacuum input states (Section \ref{subsec5b}).
Relations to other states are given in Section \ref{sec3}. 
In Section \ref{sec6} effects of 
nonperfect preparation and measurement of photon-number states are 
addressed. Finally, a sum\-mary and concluding remarks 
are given in Section \ref{sec7}. 


\section{Scheme of conditional measurement}
\label{sec2}

Splitting and mixing optical fields on beam splitters are basic 
manipulations in classical as well as in quantum optics. 
The input--output relations of a lossless beam
splitter are well known to obey  the $\rm SU(2)$ Lie algebra
\cite{Yurke4,Campos1}. In the Heisenberg picture, the photon destruction
operators of the outgoing modes, $\hat{b}_{k}$ ($k$ $\!=$ $\!1,2$), are
obtained from those of the incoming modes, $\hat{a}_{k}$, as
\begin{equation}
\hat{b}_{k} = \sum_{k'=1}^{2} T_{k,k'} \, \hat{a}_{k'},
\label{1.01}
\end{equation}
where
\begin{eqnarray}
(T_{k,k'}) = e^{i\varphi_0}
\left(\begin{array}{cc}\cos\theta
\;e^{i\varphi_{T}}&\sin\theta\;e^{i\varphi_{R}}\\
-\sin\theta\;e^{-i\varphi_{R}}&\cos\theta\;e^{-i\varphi_{T}}
\end{array}\right)
\label{1.02}
\end{eqnarray}
is a SU(2) matrix whose elements are given by the complex
transmittance $T$ and reflectance $R$ of the beam splitter,
\begin{equation}
\label{1.02a}
T = \cos\theta\;e^{i\varphi_{T}}, \quad 
R = \sin\theta\;e^{i\varphi_{R}}. 
\end{equation}
In the Schr\"{o}dinger picture, the density operator is unitarily 
transformed, whereas the photonic operators
are left unchanged. In this
case the output-state density operator $\hat{\varrho}_{\rm out}$ can be
related to the input-state density operator $\hat{\varrho}_{\rm in}$ as
\begin{equation}
\hat{\varrho}_{\rm out}
= \hat{V}^{\dagger} \hat \varrho_{\rm in}  \hat{V},
\label{1.03}
\end{equation}
where $\hat{V}$ can be given by \cite{Yurke4,Campos1}
\begin{equation}
\hat{V} =
e^{-i(\varphi_{T}-\varphi_{R}) \hat{L}_{3}}
\, e^{-2i\theta \hat{L}_{2}}
\, e^{-i(\varphi_{T}+\varphi_{R}) \hat{L}_{3}},
\label{1.03a}
\end{equation}
with
\begin{equation}
\hat{L}_{2} = \textstyle\frac{1}{2i}(\hat{a}_{1}^\dagger\hat{a}_{2}
-\hat{a}_{2}^\dagger\hat{a}_{1}), \quad
\hat{L}_{3} = \textstyle\frac{1}{2}(\hat{a}_{1}^\dagger\hat{a}_{1}
-\hat{a}_{2}^\dagger\hat{a}_{2}).
\label{1.03b}
\end{equation}
Note that $\varphi_{0}$ is a global phase factor, which may be omitted
without loss of generality, $\varphi_{0}\!=\!0$. Applying
elementary parameter-differentiation techniques \cite{Wilcox},
we can derive the operator identity
\begin{equation}
e^{-2i\theta \hat{L}_{2}}=e^{\tan\theta \, \hat{a}_{2}^\dagger\hat{a}_{1}}
\,e^{2\ln\cos\theta \, \hat{L}_3}
\,e^{-\tan\theta \, \hat{a}_{1}^\dagger\hat{a}_{2}},
\end{equation}
which [together with equation(\ref{1.02a})] enables us to
rewrite $\hat{V}^{\dagger}$, equation(\ref{1.03a}), as
\begin{equation}
\hat{V}^{\dagger} =
T^{\hat{n}_{1}}
\, e^{-R^*\hat{a}^\dagger_{2}\hat{a}_{1}}
\, e^{R\hat{a}^\dagger_{1}\hat{a}_{2}}
\,T^{-\hat{n}_{2}}\,,
\label{1.03c}
\end{equation}
where  $\hat{n}_{k}=\hat{a}^\dagger_{k}\hat{a}_{k}$.

\begin{figure}[tbh]
\centering\epsfig{figure=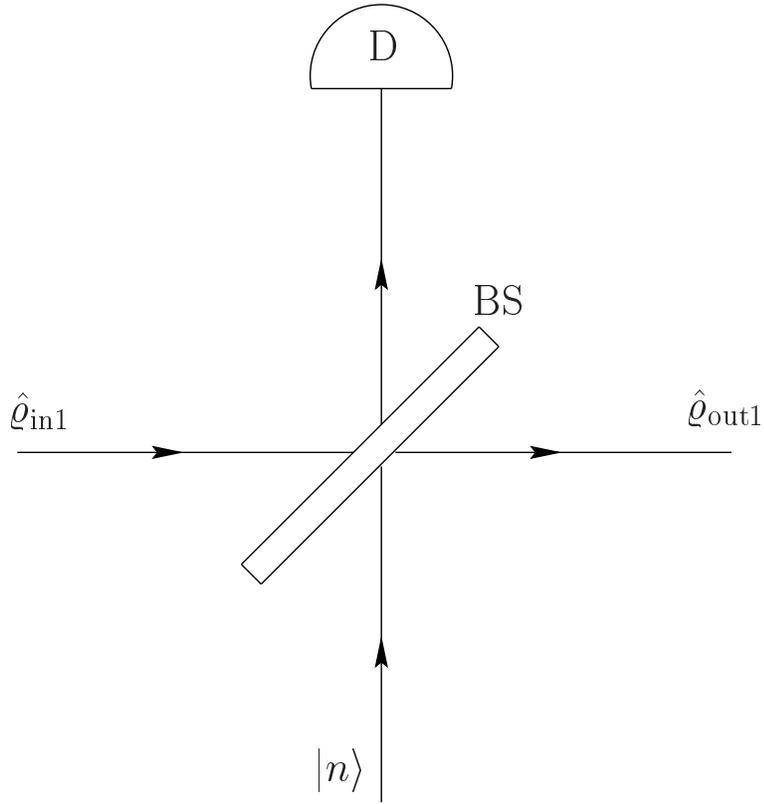,width=0.7\linewidth}
\caption{Scheme of the experimental setup. When a signal
mode prepared in a state $\hat{\varrho}_{{\rm in}1}$
is mixed (beam splitter BS) with another input mode prepared in a 
Fock state $|n\rangle$ and in one of the output channels of the 
beam splitter $m$ photons are recorded (detector D), then the quantum 
state $\hat{\varrho}_{{\rm out}1}$ of the mode in the other output 
channel ``collapses'' to either a PSJP state ($n$ $\!<$ $\!m$)
or a PAJP state ($n$ $\!>$ $\!m$).}
\label{Fig1}
\end{figure}

An outline of the experimental setup is depicted in Figure \ref{Fig1}. 
A field mode prepared in a state described by the density 
operator $\hat{\varrho}_{{\rm in}1}$ is mixed at a beam splitter 
with another mode prepared in a Fock state $|n\rangle$.
The input-state density operator can then be written as
\begin{equation}
\hat \varrho_{\rm in}(n) = \hat \varrho_{{\rm in}1}
\otimes |n\rangle_{2} \, _{2}\langle n|.
\label{1.04}
\end{equation}
Using equations (\ref{1.03}), (\ref{1.03c}) and (\ref{1.04}),
after some algebra the output-state density operator
$\hat{\varrho}_{\rm out}$ $\!\equiv$ $\!\hat{\varrho}_{\rm out}(n)$ 
can be given by
\begin{eqnarray}
\lefteqn{
\hat \varrho _{\rm out}(n) = \frac{1}{|T|^{2n}}
\sum_{l=0}^{\infty}\sum_{m=0}^{\infty}
\sum_{k=0}^{n}\sum_{j=0}^{n} (R^*)^{m+j} R^{l+k} 
}
\nonumber \\  &&  
\times 
\frac{(-1)^{l+m} }{ \sqrt{k!j!m!l!} }
\sqrt{\! {n\choose k}\!{n\choose j}\!
{n\!-\!k\!+\!m\choose m}\!{n\!-\!j\!+\!l\choose l} }  
\nonumber \\  &&  
\times \,
T^{\hat{n}_{1}}
{\hat{a}_{1}}^{m}({\hat{a}_{1}^\dagger})^{k}
\hat{\varrho}_{{\rm in}1}\hat{a}_{1}^{j}
({\hat{a}_{1}^\dagger})^{l}(T^*)^{\hat{n}_{1}}
\!\otimes \!
|n\!-\!k\!+\!m\rangle_{2} \, _{2}\langle n\!-\!j\!+\!l|.
\nonumber \\  &&  
\label{1.05}
\end{eqnarray}
   From equation (\ref{1.05}) we see that the output modes are highly 
correlated to each other in general. When the photon number of the mode 
in the second output channel is measured and $m$ photons are detected, 
then the mode in the first output channel is prepared in a quantum state 
whose density operator $\hat{\varrho}_{{\rm out}1}(n,m)$ reads as
\begin{eqnarray}
\hat{\varrho}_{{\rm out}1}(n,m)
= \frac{_{2}\langle m |\hat{\varrho}_{\rm out}(n) | m \rangle_{2}}
{{\rm Tr}_{1}\{ _{2}\langle m |
\hat{\varrho}_{\rm out}(n) | m \rangle_{2}\}} \,.
\label{1.06}
\end{eqnarray}
The probability of such an event is given by
\begin{eqnarray}
\lefteqn{
\hspace*{-2ex}
P(n,m)=
{\rm Tr}_{1}\{ _{2}\langle m |
\hat{\varrho}_{\rm out}{(n)} | m \rangle_{2} \}
}
\nonumber \\ && 
= \, \frac{|R|^{-2\nu}n!}{|T|^{2m}m!}  
\sum_{j=\mu}^{n}\sum_{k=\mu}^{n}
(-|R|^2)^{(j+k)}{m\choose j\!-\!\nu} {m\choose k\!-\!\nu}
\nonumber \\ && \hspace{2ex}\times \,
\sum_{n=\delta}^{\infty}\frac{n!|T|^{2n}}{(n\!+\!\nu)!}
{n\!+\!j\choose j}{n\!+\!k\choose k}
\langle n | \hat{\varrho}_{{\rm in}1} | n \rangle,
\label{1.07}
\end{eqnarray}
where the abbreviations
\begin{equation}
\nu = n - m, \quad
\mu=\max(0,\nu), \quad \delta=\mu-\nu
\end{equation}
have been used.
Let us now assume that the mode in the first input channel is 
prepared in a mixed state 
\begin{equation}
\hat{\varrho}_{{\rm in}1}
= \sum_{\Phi} \tilde p_{\Phi} \, | \Phi\rangle\langle \Phi  |
\label{1.08}
\end{equation}
($\sum_{\Phi}\tilde p_{\Phi}$ $\!=$ $\!1$, $0$ $\!\leq$ 
$\!\tilde p_{\Phi}$ $\!\leq$ $\!1$). 
Combining equations (\ref{1.05}) and (\ref{1.06}) and using
equation (\ref{1.08}), we find that the
mode in the first output channel is prepared in a state
\begin{equation}
\hat{\varrho}_{{\rm out}1}(n,m)
= \sum_{\Phi} \tilde p_{\Phi}| \, \Psi_{n,m} \rangle 
\big\langle \Psi_ {n,m}| ,
\label{1.09}
\end{equation}
where
\begin{equation}
|\Psi_{n,m}\rangle=
{\cal N}_{n,m}^{-1/2}
\sum_{k=\mu}^{n}\frac{(-|R|^2)^{k}}{(k\!-\!\nu)!} {n\choose k}
\hat{a}_1^{k-\nu}(\hat{a}^\dagger_1)^k T^{\hat{n}_1} |\Phi\rangle,
\label{1.10}
\end{equation}
${\cal N}_{n,m}$ being the normalization constant,
\begin{eqnarray}
\label{1.11}
\lefteqn{
\hspace*{-2ex}
{\cal N}_{n,m}
= \sum_{k=\mu}^{n}\sum_{j=\mu}^{n}\frac{(-|R|^2)^{k+j}}
{(k\!-\!\nu)!(j\!-\!\nu)!}
{n\choose k}{n\choose j}
}
\nonumber \\ && \hspace{5ex}\times \,
\langle \Phi | (T^{\ast})^{\hat{n}_1} \hat{a}_{1}^{j}
(\hat{a}^\dagger_1)^{j-\nu}\hat{a}_1^{k-\nu}(\hat{a}^\dagger_1)^{k}
 T^{\hat{n}_1} | \Phi \rangle .
\end{eqnarray}


\section{Photon-subtracted and photon-added Jacobi polynomial states}
\label{sec2a}

The properties of the conditional output state essentially depend
on whether photons are effectively subtracted ($n$ $\!<$ $\!m$)
or added  ($n$ $\!>$ $\!m$). To be more specific, from equation 
(\ref{1.10}) we obtain for $n$ $\!<$ $\!m$ the output state
\begin{eqnarray}
\lefteqn{
\hspace*{-2ex}
|\Psi_{n,m}\rangle\!=\! 
{\cal N}_{n,m}^{-1/2}
\sum_{k=0}^{n}\frac{(-|R|^2)^{k}}{(k\!+\!|\nu|)!} {n\choose k}
\hat{a}^{|\nu|} \hat{a}^k (\hat{a}^\dagger)^k  T^{\hat{n}} 
|\Phi\rangle
}
\nonumber \\ && \hspace{-2ex}\!=\!
{\cal N}_{n,m}^{-1/2}\hat{a}^{|\nu|}\Bigg\{
\sum_{k=0}^{n}\frac{(-|R|^2)^{k}k!}{(k\!+\!|\nu|)!} {n\choose k}
{\hat{n}\!+\!k\choose k}\Bigg\} T^{\hat{n}} |\Phi\rangle,
\quad
\label{1.21}
\end{eqnarray}
where the notation
\begin{eqnarray}
\label{1.21a}
\hat{a}^{k}(\hat{a}^\dagger)^k
=(\hat{n}+1)(\hat{n}+2)\cdots(\hat{n}+k) =
k!{\hat{n}\!+\!k\choose k}
\quad
\end{eqnarray}
has been introduced ($\hat{n}$ $\!\equiv$ $\!\hat{n}_{1}$). 
For $n$ $\!>$ $\!m$ we derive
[on using the relation $\hat{n}(\hat{a}^\dagger)^{l}$ $\!=$ 
$\!(\hat{a}^\dagger)^{l}$ $\!(\hat{n}$ $\!+$ $\!l)$]
\begin{eqnarray}
\lefteqn{
\hspace*{-2ex}
|\Psi_{n,m}\rangle  = 
\frac{n!(-|R|^2)^{\nu} }{m!{\cal N}_{n,m}^{1/2}}
}
\nonumber \\ && \hspace{2ex}
\times
\sum_{k=0}^{m}\frac{(-|R|^2)^{k}}{(k\!+\!\nu)!} {m\choose k}
\hat{a}^{k}(\hat{a}^\dagger)^k
(\hat{a}^\dagger)^{\nu} T^{\hat{n}} |\Phi\rangle,
\nonumber \\ && 
= \,\frac{(-|R|^2)^{\nu} n!}{m!{\cal N}_{n,m}^{1/2}}(\hat{a}^\dagger)^{\nu}
\nonumber \\ && \hspace{2ex}\times\
\Bigg\{\sum_{k=0}^{m}\frac{(-|R|^2)^{k}k!}{(k\!+\!\nu)!} {m\choose k}
{\hat{n}\!+\!\nu\!+\!k\choose k}\Bigg\} T^{\hat{n}} 
|\Phi\rangle.
\quad
\label{1.20a}
\end{eqnarray}
   From equations (\ref{1.21}) and (\ref{1.20a}) 
it can be shown (Appendix A)
that the conditional output state is of the form  
\begin{eqnarray}
|\Psi_{n,m}\rangle\sim  
\left\{
\begin{array}{ll}
\hat{a}^{|\nu|}\, 
{\rm P}_n^{(|\nu|,\hat{n}-m)}(2|T|^2\!-\!1)
T^{\hat{n}}|\Phi\rangle 
& {\rm for} \ \nu\!<\!0, \\[3ex]                 
(\hat{a}^\dagger)^{\nu}
{\rm P}_m^{(\nu,\hat{n}-m)}(2|T|^2\!-\!1)
T^{\hat{n}}|\Phi\rangle
& {\rm for} \  \nu\!>\!0,                        
\end{array}\right.
\quad
\label{jacobi}
\end{eqnarray}
where ${\rm P}_l^{(\alpha,\beta)}(z)$ is the Jacobi polynomial. 
The following procedure is seen to yield the conditional output states 
from a chosen input state $|\Phi\rangle$ $\!=$ 
$\!\sum_{k=0}^\infty\,c_k|k\rangle$. ($i$) Replace the Fock-expansion 
coefficients $c_k$ with 
$c'_k$ $\!\sim$ $\!{\rm P}_{n-\mu}^{(|\nu|,k-m)}$ $\!(2|T|^2$ $\!-$ 
$\!1)$ $\!T^{k}c_k$ to obtain a state $|\Phi'\rangle$.
($ii$) Subtract photons from the state $|\Phi'\rangle$ by repeated 
application of the photon-destru\-ction operator to it or add photons 
to state $|\Phi'\rangle$ by repeated application of the photon-creation 
operator to it. In what follows we will refer to the states 
$|\Phi'\rangle$ as Jacobi polynomial (JP) states (in analogy to 
Hermite polynomial and Laguerre polynomial states \cite{Bergou,Fu}).
Note that for typical classes of states the input state $|\Phi\rangle$
and the state $|\Psi\rangle$ $\!\sim$ $\!T^{\hat{n}}|\Phi\rangle$ belong 
to the same class of states \cite{Dakna3}. We see that
the conditional output states $|\Psi_{n,m}\rangle$ produced
in the scheme can be regarded as photon-subtracted Jacobi polynomial 
(PSJP) states ($n$ $\!<$ $\!m$) and photon-added Jacobi polynomial 
(PAJP) states ($n$ $\!>$ $\!m$).
It should be pointed out that the PSJP states and the 
PAJP states are essentially different from each other in general, 
because of $[\hat{a},\hat{a}^\dagger]$ $\!\neq$ $\!0$. Clearly, when the
input state is a Fock state, $|\Phi\rangle$ $\!=$ $\!|k\rangle$, 
then the conditional output states are the Fock states 
$|k$ $\!+$ $\!n$ $\!-$ $\!m\rangle$.
Let us mention that when $m$ $\!=$ $\!n$ (i.e., $\nu$ $\!=$ $\!0$), then
\begin{eqnarray}
|\Psi_{n,n}\rangle={\cal N}_{n,n}^{-1/2}
{\rm P}_{n}^{(0,\hat{n}-n)}(2|T|^2\!-\!1)T^{\hat{n}}
|\Phi\rangle\,.
\end{eqnarray}


\section{PSJP and PAJP coherent states}
\label{subsec5a}

To treat the states in a unified way, let us return to equation 
(\ref{1.10}) and first consider Glauber coherent input states 
\begin{equation}
|\Phi\rangle \equiv |\beta\rangle 
= e^{-|\beta|^2/2}\sum_{k=0}^{\infty}\frac{\beta^k}
{\sqrt{k!}}\,|k\rangle,
\label{2.1}
\end{equation}
with $\beta$ $\!=$ $\!|\beta|e^{i\varphi_\beta}$. 
Equation (\ref{1.10}) then reads
\begin{equation}
|\Psi_{n,m}\rangle
= \frac{1}{\sqrt{{\cal N}'_{n,m}}}
\sum_{k=\mu}^{n}\frac{(-|R|^2)^k}{(k-\nu)!} {n\choose k}
\,\hat{a}^{k\!-\!\nu}(\hat{a}^\dagger)^k |\beta'\rangle,
\label{2.2}
\end{equation}
where $\beta'$ $\!=$ $\!T \beta$ and 
${\cal N}'_{n,m}$ $\!=$ $\!e^{|\beta'R|^2}{\cal N}_{n,m}$. 
Applying standard operator ordering techniques \cite{Vogel2},
we may write
\begin{eqnarray}
\hat{a}^{m}(\hat{a}^\dagger)^n=\sum_{l=0}^{\min\{m,n\}}{m\choose l}
\frac{n!}{(n-l)!}
(\hat{a}^\dagger)^{n-l}\hat{a}^{m-l},
\label{2.4}
\end{eqnarray}
and hence
\begin{eqnarray}
\lefteqn{
|\Psi_{n,m}\rangle=\frac{{\beta'}^{-\nu}}{\sqrt{{\cal N}'_{n,m}}}
\sum_{k=\mu}^{n}(-|R|^2)^k {n\choose k}
}
\nonumber \\ && \hspace{10ex} \times \,
\sum_{l=\mu}^{k}\frac{1}{(k-\nu)!}{k \choose l}
(\beta'\hat{a}^\dagger)^{l}|\beta'\rangle,
\label{2.3}
\end{eqnarray}
from which the Fock-state expansion of $|\Psi_{n,m}\rangle$ can easily
be obtained to be
\begin{eqnarray}
\lefteqn{
\hspace{-2ex}
|\Psi_{n,m}\rangle=\frac{e^{-|\beta'|^2/2}}{\sqrt{{\cal N}'_{n,m}}}
\sum_{k=\mu}^{n}(-|R|^2)^k {n\choose k}
}
\nonumber \\ && \hspace{2ex} \times \,
\,\sum_{l=\mu}^{k}{k\choose l}\frac{{\beta'}^{(l-\nu)}}{(l-\nu)!}
\sum_{p=0}^{\infty}\frac{(\beta')^{p}}{p!}\sqrt{(p+l)!}
\; |p+l\rangle
\quad
\label{2.7}
\end{eqnarray}
(for the photon-number statistics, see Appendix B).
Using the identities \cite{Prudnikov1}
\begin{eqnarray}
\sum_{l=0}^{n}\frac{x^l}{\Gamma(l+\nu)}{n\choose l}
=\frac{n!}{\Gamma(n+\nu)}{\rm L}_n^{\nu-1}(-x)
\label{2.4a}
\end{eqnarray}
and
\begin{equation}
\sum_{l=0}^{n}\frac{t^l}{\Gamma(\alpha\!+\!l\!+\!1)}
\frac{{\rm L}_l^\alpha(x)}{(n\!-\!l)!}
\!=\!\frac{(1+t)^n{\rm L}_n^\alpha\left[tx/(1\!+\!t)\right]}
{\Gamma(\alpha\!+\!n\!+\!1)}
\label{2.4aa}
\end{equation}
[${\rm L}_n^{\alpha}(z)$ is the  associated (or generalized) Laguerre 
polynomial \cite{Prudnikov1}], we may give equation (\ref{2.3}) 
in the more compact form of
\begin{eqnarray}
\lefteqn{
\hspace*{-2ex}
|\Psi_{n,m}\rangle
= \frac{|T|^{2n}n!}{{\beta'}^{\nu}\sqrt{{\cal N}'_{n,m}}}
\frac{\left[-|R|^2/|T|^2\right]^{\mu}}{(n+\delta)!}
}
\nonumber \\ && \hspace{12ex} \times \,
\,{\rm L}_{n-\mu}^{|\nu|}
\!\left(\frac{|R|^2}{|T|^2}\beta'\hat{a}^\dagger \right)
(\beta'\hat{a}^\dagger)^{\mu}
 |\beta'\rangle.
\label{2.4b}
\end{eqnarray}
In a similar way it follows that
\begin{eqnarray}
\lefteqn{
\hspace*{-2ex}
{\cal N}'_{n,m}=\frac{1}{|\beta'|^{2\nu}}\sum_{k=\mu}^{n}
\sum_{j=\mu}^{n}(-|R|^2)^{k+j}{n\choose k}{n\choose j}
}
\nonumber \\ && \hspace{6ex} \times \,
\sum_{l=\mu}^{k}\sum_{l'=\mu}^{j}{k \choose l}{j\choose l'}
\frac{{\beta'}^{l} (\beta'^\ast)^{l'}
\chi_{l',l}^{(1)}(\beta')}{(l-\nu))!(l'-\nu)!}\,,
\label{2.5}
\end{eqnarray}
where 
\begin{eqnarray}
\chi_{l,k}^{(1)}(\alpha)
=\left\{
\begin{array}{ll}
k!{\alpha}^{l-k}{\rm L}_{k}^{l-k}(-|\alpha|^2) 
&  {\rm for} \ l\ge k, \\ [.5ex] 
l!(\alpha^\ast)^{k-l}{\rm L}_{l}^{k-l}(-|\alpha|^2)
&  {\rm for} \ l < k.
\end{array}
\right.
\label{2.6}
\end{eqnarray}
  From equations (\ref{2.3}) or (\ref{2.4b}) we find that 
PSJP and PAJP coherent states are finite superpositions of
photon-added coherent states. In particular for $n$ $\!<$ $\!m$ 
equation (\ref{2.4b}) reduces to
\begin{equation}
\label{2.6a}
|\Psi_{n,m}\rangle
\sim {\rm L}_{n}^{|\nu|}
\!\left(\frac{|R|^2}{|T|^2}\beta'\hat{a}^\dagger \right) |\beta'\rangle,
\end{equation}
from which we see that when $|\beta'|$ $\!\ll$ $\!1$ and 
$\gamma$ $\!=$ $\!(|R|^2/|T|^2)\beta'$ finite, then
the PSJP state $|\Psi_{n,m}\rangle$ is (approximately) a superposition
of $n$ Fock states, because of $|\beta'\rangle$ $\!\approx$ $\!|0\rangle$.

\begin{figure}[tbh]
\centering\epsfig{figure=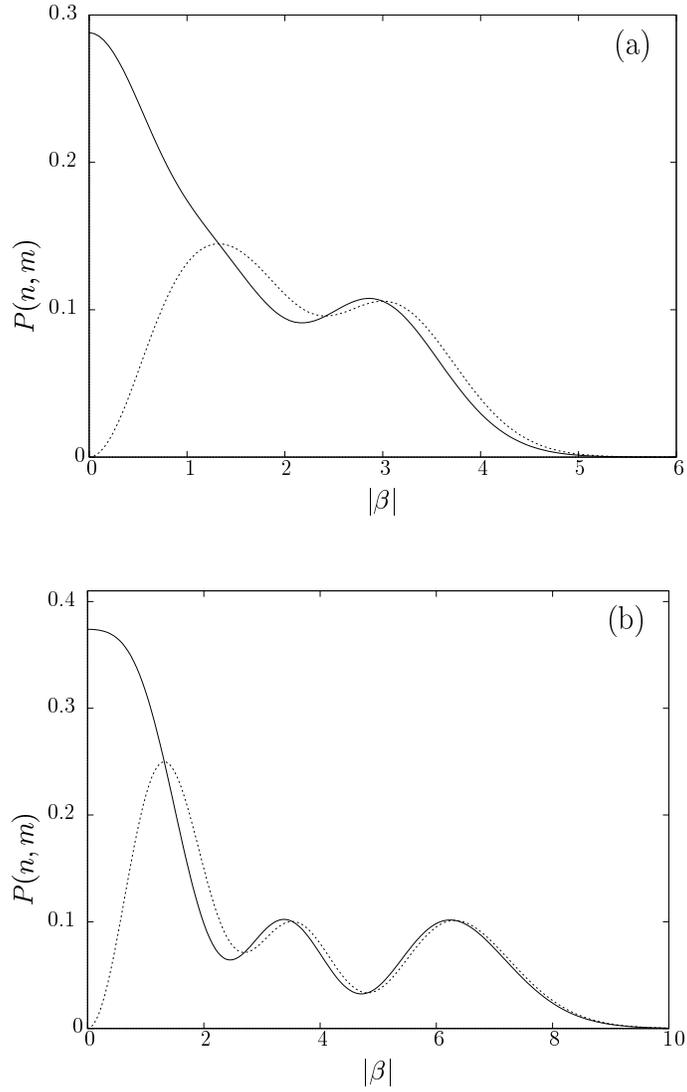,width=0.7\linewidth}
\caption{
The probability of producing ($n$ $\!=$ $\!2$, $m$ $\!=$ $\!3$) 
PSJP coherent states (dashed line) and ($n$ $\!=$ $\!3$, $m$ $\!=$ $\!2$) 
PAJP coherent states (solid line) is shown as a function 
of $|\beta|$ for two values of the beam-splitter transmittance 
[(a) $|T|^2$ $\!=$ $\!0.4$; (b) $|T|^2$ $\!=$ $\!0.81$].}
\label{Fig2}
\end{figure}

The probability of producing PSJP and PAJP coherent states can
be obtained from equation (\ref{1.07}). After some calculation
we derive (Appendix C)
\begin{eqnarray}
\lefteqn{
\hspace{-2ex}
P(n,m)=e^{-|R|^2|\beta^2|}\frac{|R|^{-2\nu}n!}{|T|^{2m}m!}
}
\nonumber \\ && \hspace{0ex}\times \,
\sum_{k=\mu}^{n}\sum_{j=\mu}^{n}(-|R|^2)^{k+j}
{m\choose k\!-\!\nu}{m\choose j\!-\!\nu}\chi^{(2)}_{k,j}(\beta',\nu),
\quad
\label{2.18}
\end{eqnarray}
where
\begin{eqnarray}
\lefteqn{
\chi_{k,j}^{(2)}(\alpha,\nu)=
}
\nonumber \\ && \hspace{0ex} \,
\left\{
\begin{array}{ll}
\displaystyle\sum_{l=0}^{k}{k\choose l}\frac{(j-\nu)!)}{l!j!}
{\rm L}_{j-\nu}^{l+|\nu|}(-|\alpha|^2)|\alpha|^{2l}  
& {\rm for} \ \nu\ge 0, \\ [4ex] 
\displaystyle\sum_{l=0}^{k}{k\choose l}\frac{(j+l)!)}{l!j!}
{\rm L}_{j+l}^{-l+|\nu|}(-|\alpha|^2)|\alpha|^{2|\nu|}
& {\rm for} \ \nu<0 .
\end{array}
\right.
\quad
\label{2.19}
\end{eqnarray}
In Figure \ref{Fig2} examples of $P(n,m)$ are plotted for two absolute 
values of the beam-splitter transmittance. We see that $P(n,m)$ 
is an oscillating function of the absolute value of the coherent input 
amplitude, which is due to the interference of the incoming fields at the 
beam splitter. As expected, for $n$ $\!<$ $\!m$ the probability $P(n,m)$ 
goes to zero as the coherent amplitude does.

\begin{figure}[tbh]
\centering\epsfig{figure=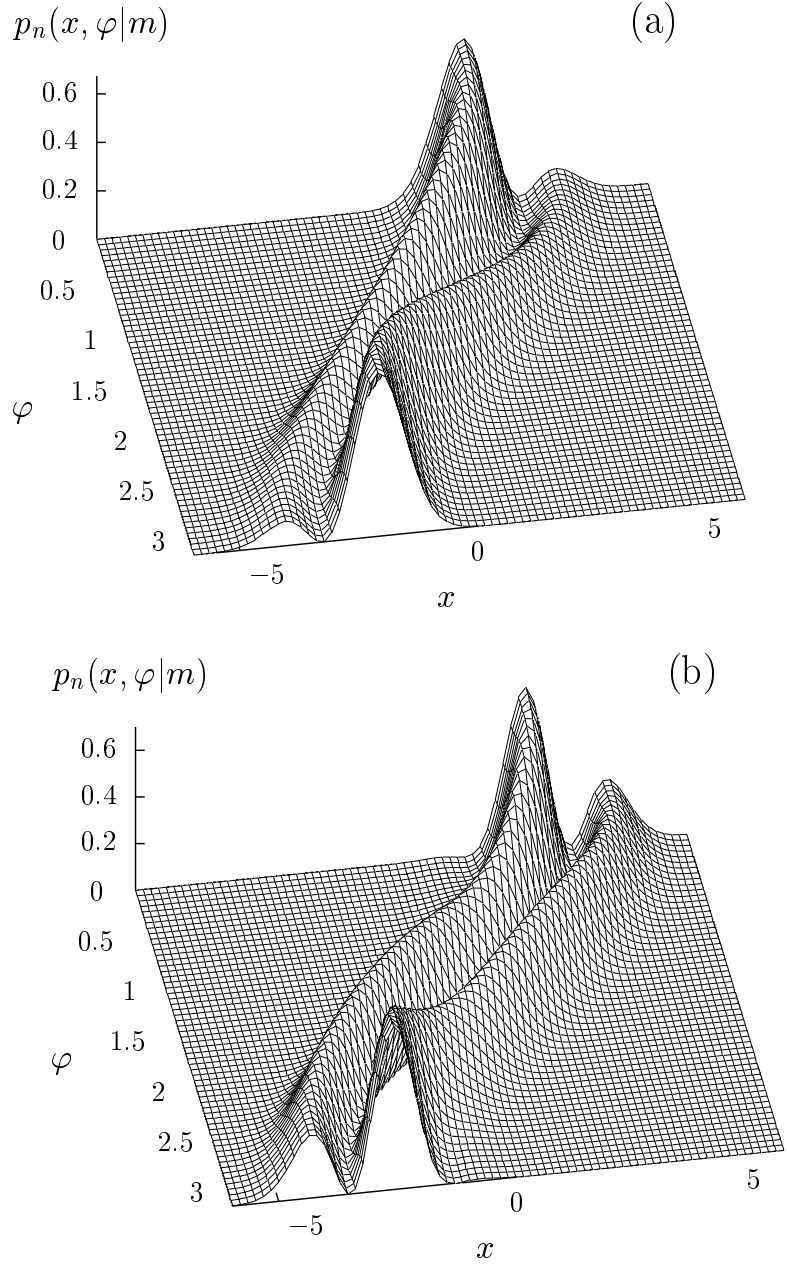,width=0.7\linewidth}
\caption{The quadrature-component distributions of 
(a) PSJP and (b) PAJP coherent states for $\beta'$ $\!=$ $\! 2.07$
($|\beta|$ $\!=$ $\! 2.3$, $|T|^2$ $\!=$ $\!0.81$)
[(a) $n$ $\!=$ $\!2$, $m$ $\!=$ $\!3$,
(b) $n$ $\!=$ $\!3$, $m$ $\!=$ $\!2$]. }
\label{Fig3}
\end{figure}

The superposition of photon-added coherent states, equation (\ref{2.3}), 
gives rise to strong quantum interference, as it can be seen from the 
quadrature distribution 
\begin{equation}
p_{n}(x,\varphi|m)=|\langle x,\varphi|\Psi_{n,m}\rangle|^2.
\label{2.7a}
\end{equation}
Using the Fock-state expansions (\ref{2.7}) and \cite{Vogel2}
\begin{equation}
|x,\varphi\rangle
=(\pi)^{-1/4} e^{-x^2/2}
\sum_{k=0}^{\infty}\frac{e^{ik\varphi}}
{\sqrt{2^k k!}} \, {\rm H}_k(x) |k \rangle
\label{2.8}
\end{equation}
[${\rm H}_k(x)$, Hermite polynomial] 
of the states $|\Psi_{n,m}\rangle$
and $|x,\varphi\rangle$, respectively, 
and recalling the identity \cite{Prudnikov1}
\begin{eqnarray}
\sum_{k=0}^{\infty}\frac{z^k}{k!}\,{\rm H}_{k+n}(x)
= \, \exp\!\left(2xz-z^2\right)
{\rm H}_{n}(x-z),
\label{2.9}
\end{eqnarray}
we find that 
\begin{eqnarray}
\lefteqn{
\hspace{-2ex}
p_{n}(x,\varphi|m)
= \frac{|\beta'|^{-2\nu}}{{\cal N}'_{n,m}\sqrt{\pi}}
}
\nonumber \\ &&  \hspace{-2ex}
\times  
\exp\!\left\{\!-\!
\left[
x\!-\!\sqrt{2}|\beta'|\cos(\varphi\!-\!\varphi_{\beta'})
\right]^2
\right\}
\Bigg |\sum_{k=\mu}^{n}(-|R|^2)^k 
\nonumber \\ && \hspace{-1ex}
\times 
{n\choose k}\sum_{l=\mu}^{k}{k\choose l}\frac{(2^{-\frac{1}{2}}{\beta'}^\ast
 e^{i\varphi})^{l}}
{(l-\nu)!}{\rm H}_{l}\!\left(
x \!- \!2^{-\frac{1}{2}}{\beta'}^\ast e^{i\varphi}\right)\Bigg |^2\!\!,
\quad
\label{2.10}
\end{eqnarray}
where $\varphi_{\beta'}\!=\!\varphi_{\beta}+\varphi_T$.
Examples of $p_{n}(x,\varphi|m)$ are plotted in 
Figure \ref{Fig3} for $\beta'$ $\!=$ $\!2.07$ and 
(a) $n=2$, $m=3$ [$P(n,m)\approx 9.7\%$] and 
(b) $n=3$, $m=2$  [$P(n,m)\approx 6.7\%$]. From the figure it is clearly 
seen that the states are extremely non-Gaussian squeezed coherent states
owing to quantum interference.

\begin{figure}[tbh]
\centering\epsfig{figure=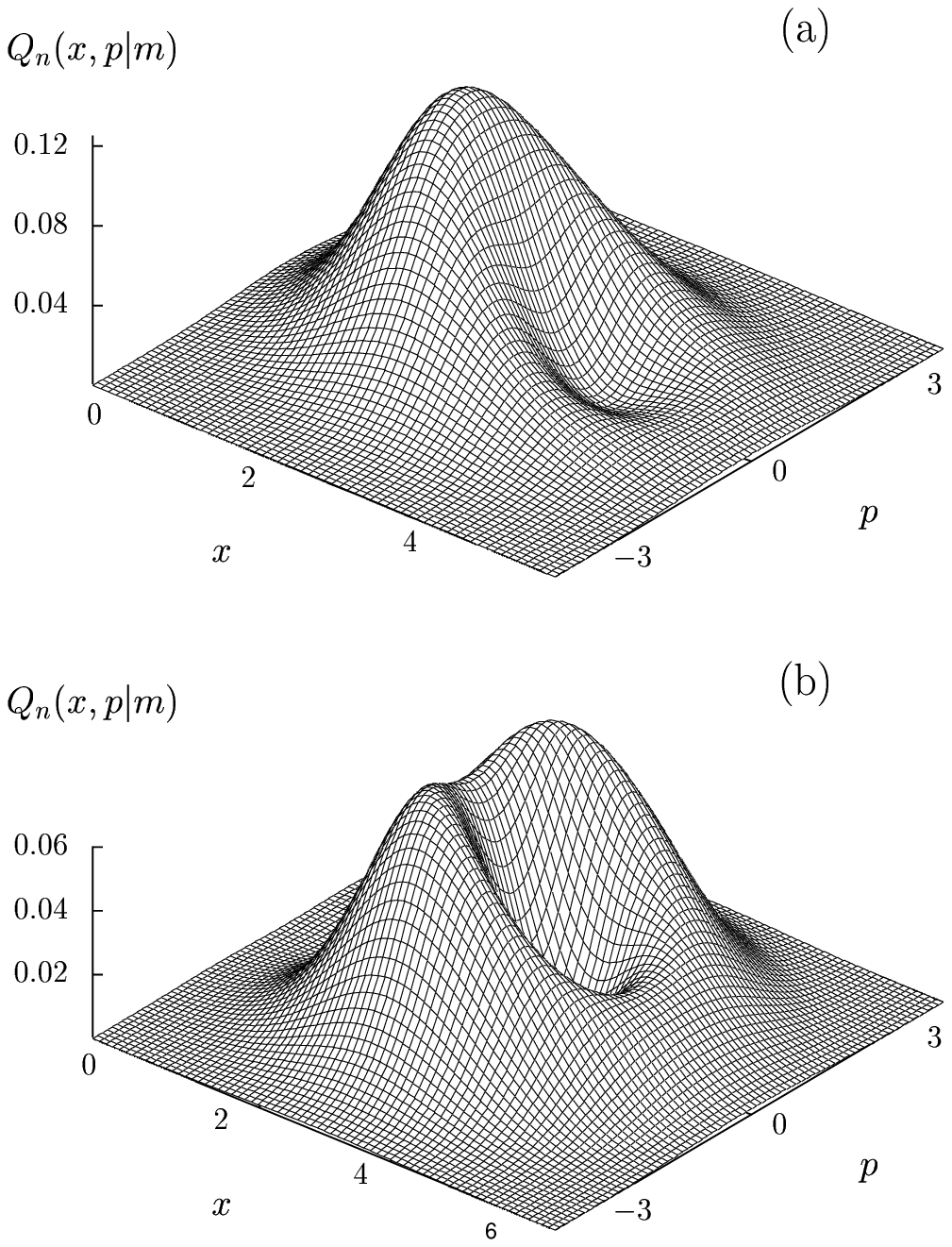,width=0.7\linewidth}
\caption{The Husimi function of 
(a) PSJP and (b) PAJP coherent states for $\beta'$ $\!=$ $\! 2.07$
($|\beta|$ $\!=$ $\! 2.3$, $|T|^2$ $\!=$ $\!0.81$)
[(a) $n$ $\!=$ $\!2$, $m$ $\!=$ $\!3$,
(b) $n$ $\!=$ $\!3$, $m$ $\!=$ $\!2$]. }
\label{Fig4}
\end{figure}

Next let us consider the Husimi function 
\begin{equation}
Q_{n}(x,p|m)= \frac{1}{2\pi} \, |\langle\alpha|\Psi_{n,m}\rangle|^2 ,
\label{2.11}
\end{equation}
with $|\alpha \rangle$ being a coherent state and
$\alpha$ $\!=$ $\!2^{-1/2}(x\!+\!ip)$. Expanding $|\alpha\rangle$
and $|\Psi_{n,m}\rangle$ in the Fock basis, on applying equations
(\ref{2.1}) and (\ref{2.7}) respectively, we derive
\begin{eqnarray}
\lefteqn{
Q_{n}(x,p|m)
=\frac{\exp\!\left[-|\alpha\!-\!\beta'|^2\right]}
{2\pi|\beta'|^{2\nu}{\cal N}'_{n,m}}
}
\nonumber \\ && \hspace{4ex}\times 
\left|\sum_{k=\mu}^{n}(-|R|^2)^k {n\choose k}
\sum_{l=\mu}^{k}{k\choose l}\frac{(\alpha{\beta'}^\ast)^{l} }
{(l\!-\!\nu)!}\right |^2\!.
\label{2.12}
\end{eqnarray}
We again use the relations (\ref{2.4a}) and (\ref{2.4aa}) and
rewrite equation (\ref{2.12}) as

\begin{eqnarray}
\lefteqn{
\hspace*{-2ex}
Q_{n}(x,p|m)=\frac{|T|^{4n}(n!)^2}{2\pi{\cal N}'_{n,m}
|\beta'|^{2\nu}} \frac{(-|RT\alpha\beta'|)^{2\mu}}
{(n+\delta)!}
}
\nonumber \\ && \hspace{12ex}\times \,
e^{-|\alpha-\beta'|^2}
\left|{\rm L}_{n-\mu}^{|\nu|}
\!\left(\frac{|R|^2\alpha{\beta'}^\ast}{|T|^2} \right)\right|^2\!.
\end{eqnarray}
Figure \ref{Fig4} shows plots of the Husimi function 
for $\beta'$ $\!=$ $\!2.07$ and
(a) $n$ $\!=$ $\!2$, $m$ $\!=$ $\!3$ and (b) \mbox{$n$ $\!=$ $\!3$}, 
$m$ $\!=$ $\!2$. It should be pointed out that
(the square root of) the difference between the height of
the Husimi function and $1/(2\pi)$ can be regarded, in a sense,
as a measure of the degree of nonclassicality 
\cite{Dodonov1,Wuensche4}.
For the most classical states, i.e., for the Glauber coherent states 
and only for them, the Husimi function attains the maximum height of 
$1/(2\pi)\approx 0.16$. Thus from Figures \ref{Fig4}(a) and 
\ref{Fig4}(b) we see that the PSJP coherent state
($n$ $\!<$ $\!m$) is more classical than the
PAJP coherent state ($n$ $\!>$ $\!m$)
-- a behaviour that is observed for a wide range of values of $|\beta'|$.

\begin{figure}[tbh]
\centering\epsfig{figure=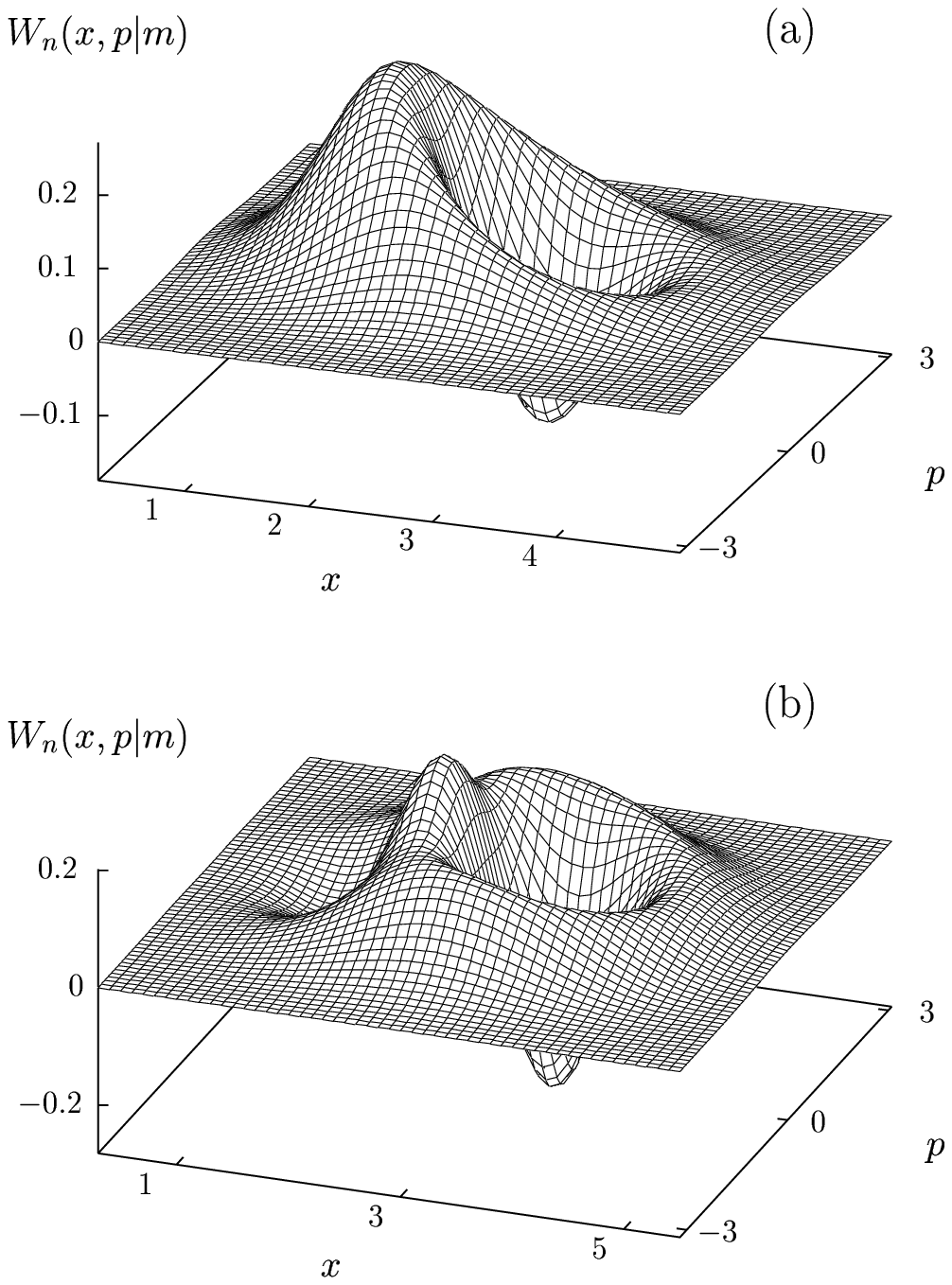,width=0.7\linewidth}
\caption{The Wigner function of 
(a) PSJP and (b) PAJP coherent states for $\beta'$ $\!=$ $\! 2.07$
($|\beta|$ $\!=$ $\! 2.3$, $|T|^2$ $\!=$ $\!0.81$)
[(a) $n$ $\!=$ $\!2$, $m$ $\!=$ $\!3$,
(b) $n$ $\!=$ $\!3$, $m$ $\!=$ $\!2$]. }
\label{Fig5}
\end{figure}

Among the phase-space functions that have been shown to be inferable 
from measurable data, the Wigner function 
\begin{equation}
\label{2.13}
W_{n}(x,p|m) 
= \frac{1}{\pi}\int\limits_{-\infty}^{+\infty} {d}y \, e^{2ipy}
\langle x\!-\!y|\Psi_{n,m}\rangle\langle\Psi_{n,m}|x\!+\!y\rangle
\end{equation}
($|x\rangle$ $\!=$ $\!|x,\varphi\rangle|_{\varphi=0}$) 
reflects quantum features most distinctly.  
In order to calculate it, we again use the Fock-state expansions 
(\ref{2.7}) and (\ref{2.8}) [together with equation (\ref{2.9})].
After some calculation we derive 
\begin{eqnarray}
W_{n}(x,p|m)=
\lefteqn{
\frac{e^{-\left[\!x^2+
{\textstyle\frac{1}{2}} \left(\beta'\!+\!\beta'^{\ast}\right)^2
\!-\sqrt{2}x(\beta'\!+\!\beta'^\ast)
\right]}}
{\pi{\cal N}'_{n,m}|\beta'|^{2\nu}}
}
\nonumber \\ && \hspace{-14ex} \times \,
\sum_{k=\mu}^{n}
\sum_{j=\mu}^{n}(-|R|^2)^{k+j}
{n\choose k}{n\choose j}
\sum_{l=\mu}^{k}\sum_{l'=\mu}^{j}{k \choose l}{j\choose l'}
\nonumber \\ && \hspace{-14ex}\times \,
\frac{(2^{-1/2}\beta')^{l}(2^{-1/2}{\beta'}^\ast)^{l'}}
{(l-\nu)!(l'-\nu)!}
\!\!\int\limits_{-\infty}^{+\infty} \!\!{d}y \,
e^{-y^2+y[2ip+\sqrt{2}({\beta'}^\ast-\beta')]}
\nonumber \\ && \hspace{-8ex}\times \,
{\rm H}_{l}(x\!-\!y\!-\!2^{-1/2}\beta')
{\rm H}_{l'}(x\!+\!y\!-\!2^{-1/2}{\beta'}^\ast).
\label{2.14}
\end{eqnarray}
We then use the integral identity \cite{Prudnikov1} 
\begin{eqnarray}
\lefteqn{
\int\limits_{-\infty}^{+\infty} {d}x \,
e^{-x^2}{\rm H}_{m}(x\!+\!a){\rm H}_{n}(x\!+\!b)
}
\nonumber \\ &&\hspace{5ex} =\,
\frac{\sqrt{\pi}2^n m!}{b^{m-n}}{\rm L}_{m}^{n-m}(-2ab)
\quad (m\le n),
\label{2.15}
\end{eqnarray}
and perform the $y$-integration to obtain
\begin{eqnarray}
\hspace*{-6ex}
W_{n}(x,p|m)=
\lefteqn{
\frac{e^{-|x+ip-\sqrt{2}\beta'|^2}}{|\beta'|^{2\nu}\pi{\cal N}'_{n,m}}
\sum_{k=\mu}^{n}\sum_{j=\mu}^{n}
\bigg\{
(-|R|^2)^{k+j}
}
\nonumber \\ &&\hspace{-15ex}\times \,
{n\choose k}{n\choose j}
\sum_{l=\mu}^{k}\sum_{l'=\mu}^{j} {k \choose l}{j\choose l'}
\frac{(\beta')^{l}(\beta'^\ast)^{l'}}{(l-\nu))!(l'-\nu)!}
\nonumber \\ &&\hspace{2ex}\times \,
\chi_{l',l}^{(3)}\!\left[\sqrt{2}(x\!+\!ip\!-\!2^{-1/2}\beta')\right]
\bigg\},
\end{eqnarray}
where
\begin{eqnarray}
\chi_{l,k}^{(3)}(\alpha)=\left\{
\begin{array}{ll}
(-1)^k k!{\alpha}^{l-k}{\rm L}_{k}^{l-k}(|\alpha|^2) 
& {\rm for} \ l\ge k, \\[.5ex] 
(-1)^l l!(\alpha^*)^{k-l}{\rm L}_{l}^{k-l}(|\alpha|^2)
& {\rm for} \ l<k .
\end{array}
\right.
\label{2.16}
\end{eqnarray}
In Figure \ref{Fig5} plots of the Wigner function $W_{n}(x,p|m)$ are shown
for $\beta'$ $\!=$ $\!2.07$ and (a) $n$ $\!=$ $\!2$, $m$ $\!=$ $\!3$
and (b) $n$ $\!=$ $\!3$, $m$ $\!=$ $\!2$.
We see that $W_{n}(x,p|m)$ for $n$ $\!>$ $m$ is more structurized
owing to quantum interference and stronger negative than 
for $n$ $\!<$ $m$, which again reveals that the PAJP coherent
state is more nonclassical than the PSJP coherent state.


\section{PSJP and PAJP squeezed vacuum states}
\label{subsec5b}

Let us briefly comment on PSJP and PAJP squeezed vacuum states that
are realized when the input state is a squeezed vacuum
$\hat{S}(\xi)|0\rangle$, with $\hat{S}(\xi)$ $\!=$ 
$\!\exp\{-\frac{1}{2} [\xi(\hat{a}^{\dagger})^{2}$
$\!-$ $\!\xi^{\ast}\hat{a}^{2}]\}$ being the squeeze operator.
Hence we may write 
\begin{equation}
|\Phi\rangle 
\equiv  \hat{S}(\xi)| 0 \rangle
= (1-|\kappa|^{2})^{\frac{1}{4}}\sum_{k=0}^{\infty}
\frac{[(2k)!]^{1/2}}{2^{k}k!}\,
\kappa^{k}|2k\rangle ,
\label{3.1}
\end{equation}
where the notation $\kappa$ $\!\equiv$ $\!\kappa(\xi)$ 
$\!=$ $\!-e^{i\varphi_{\xi}}\tanh|\xi|$ has been introduced
($\xi$ $\!=$ $\!|\xi|e^{i\varphi_{\xi}}$). 
According to equation(\ref{1.10}), the conditional output states 
are then given by 
\begin{equation}
|\Psi_{n,m}\rangle 
= \frac{1}{\sqrt{{\cal N}'_{n,m}}}
\sum_{k=\mu}^{n}\frac{(-|R|^2)^k}{(k-\nu)!} {n\choose k}
\hat{a}_1^{k-\nu}(\hat{a}^\dagger_1)^k
\hat{S}(\xi')| 0 \rangle ,
\label{3.2}
\end{equation}
with ${\cal N}'_{n,m}\!=
\![(1-|\kappa'|^2)/(1-|\kappa|^2)]^{1/2}\!{\cal N}_{n,m}$. 
In the photon-number basis $|\Psi_{n,m}\rangle$ reads as
\begin{eqnarray}
\lefteqn{
\hspace*{-3ex}
|\Psi_{n,m}\rangle
=\frac{(1-|\kappa'|^{2})^{\frac{1}{4}}}{\sqrt{{\cal N}'_{n,m}}}
\sum_{k=\mu}^{n}\frac{(-|R|^2)^k}{(k-\nu)!}{n\choose k}
}
\nonumber \\ && \hspace{-3ex}\times 
\!\sum_{p=\mu}^{\infty}
\frac{(p-\nu+k)!
{\textstyle\frac{1}{2}}\!\left[1\!+\!(-1)^{p-\nu}\right]
}
{\Gamma\!\left[{\textstyle\frac{1}{2}}(p\!-\!\nu)\!+\!1\right]\sqrt{p!}} 
\left({\textstyle\frac{1}{2}}\kappa'\right)^{(p-\nu)/2}  |p\rangle,
\label{3.3}
\end{eqnarray}
with $\kappa'$ $\!=$ $\!T^2\kappa$. From equation (\ref{3.3}) we easily 
see that when the difference between the number $n$ of photons in the 
second input channel of the beam splitter and the number $m$ of photons 
detected in the second output channel, i.e., the parameter
$\nu$ $\!=$ $\!n$ $\!-$ $\!m$, is even (odd),
then the mode in the first output channel 
is prepared in a PSJP or PAJP squeezed vacuum state 
$|\Psi_{n,m}\rangle$ that contains
only Fock states with even (odd) numbers
of photons.  
Similarly to ordinary photon-subtracted and photon-added squeezed
vacuum states \cite{Dakna1,Dakna2}, it can be shown that 
the PSJP and PAJP squeezed vacuum states are Schr\"{o}dinger-cat-like
states. In particular, from equation (\ref{3.3}) it can be found that 
$|\Psi_{n,m}\rangle$ can be given by a superposition of two
mesoscopically distinguishable states,
\begin{eqnarray}
|\Psi_{n,m}\rangle 
\sim
|\Psi^{(+)}_{n,m}\rangle + |\Psi^{(-)}_{n,m}\rangle ,
\label{3.4}
\end{eqnarray}
where 
\begin{equation}
|\Psi ^{(\pm)}_{n,m}\rangle
=\frac{1}{\sqrt{{{\cal N}'^{(\pm)}_{n,m}}}}
\sum_{p=\mu}^{\infty}
C^{(\pm)}_{n,m,p}(\kappa') \, |p\rangle ,
\label{3.5}
\end{equation}
with
\begin{eqnarray}
\lefteqn{
C^{(\pm)}_{n,m,p}(\kappa')=
\sum_{k=\mu}^{n}\frac{(-|R|^2)^k}{(k-\nu)!}{n\choose k}
}
\nonumber \\ && \hspace{8ex}\times \,
\frac{(p-\nu+k)!}
{\Gamma\!\left[{\textstyle\frac{1}{2}}(p\!-\!\nu)\!+\!1\right] \sqrt{p!}}
\left(\pm\sqrt{\textstyle\frac{1}{2}}\kappa'\right)^{p-\nu}\!\!\!.
\quad
\label{3.6}
\end{eqnarray}
A detailed analysis can be given in a way similar to that in 
\cite{Dakna1,Dakna2}. We therefore renounce the somewhat lengthy
calculations here.


\section{Relations to other states}
\label{sec3}

   From equation (\ref{jacobi}) and the property of the Jacobi 
polynomials that 
\begin{equation}
\label{1.00}
{\rm P}^{(\alpha,\beta)}_l(1) 
=  (-1)^l{\rm P}^{(\alpha,\beta)}_l(-1) = 
{{l\!+\!\alpha\choose l}}
\end{equation}  
we see that for sufficiently small values of $|T|$ 
($|T|$ $\!\to$ $\!0$) or values of $|T|$ close to unity 
($|T|$ $\!\to$ $\!1$) equation (\ref{jacobi}) approximately reduces to
\begin{eqnarray}
|\Psi_{n,m}\rangle
\sim  \left\{
\begin{array}{ll}
\hat{a}^{|\nu|} T^{\hat{n}} |\Phi\rangle 
& {\rm for} \  \nu = n - m < 0, \\[2ex] 
(\hat{a}^\dagger)^{\nu} T^{\hat{n}} |\Phi\rangle
& {\rm for} \  \nu = n - m > 0.
\end{array}
\right.
\label{jacobi2}
\end{eqnarray}
As expected, in these limiting cases the produced conditional states 
$|\Psi_{n,m}\rangle$ reduce to ordinary photon-subtracted 
($\nu$ $\!<$ $\!0$) or photon-added ($\nu$ $\!>$ $\!0$) states.

Further, equation (\ref{jacobi}) reveals that
when the second input mode is in the vacuum state and in the
second output channels a nonvanishing number of photons is
detected ($n$ $\!=$ $\!0$, $m$ $\!>$ $\!0$), then 
usual photon subtraction is observed, 
\begin{equation}
\label{1.12a}
|\Psi_{0,m}\rangle=(1/m!)
{\cal N}_{0,m}^{-1/2}
\, \hat{a}^{m} T^{\hat{n}} |\Phi\rangle ,
\end{equation}
independently of the value of $T$. 
Similarly, when a Fock state $|n\rangle$ is fed into the second input 
channel and a zero-photon conditional measurement is performed in the
second output channel ($n$ $\!>$ $\!0$, $m$ $\!=$ $\!0$), then 
the mode in the first output channel is prepared in 
an ordinary photon-added state
\begin{equation}
\label{1.13a}
|\Psi_{n,0}\rangle=(-|R|^2)^{n}
{\cal N}_{n,0}^{-1/2}
\, (\hat{a}^\dagger)^{n}\,T^{\hat{n}} |\Phi\rangle .
\end{equation}
Photon-subtracted states and photon-added states of the
type given in equations (\ref{1.12a}) and (\ref{1.13a}), respectively,
have been studied by several authors  
\cite{Dakna1,Dakna2,Dakna3,Ban1,Jones1,Agarwal2,Zhang1,Dodonov1}. 


\subsection{Displaced Fock states}
\label{subsec3.2}

Since the coherent states $|\beta\rangle$ $\!=$ $\!\hat{D}(\beta)|0\rangle$ 
[$\hat{D}(\beta)$ $\!=\!$ $\!\exp(\beta\hat{a}^{\dagger}$ $\!-$ 
$\!\beta^\ast\hat{a} )$] are the eigenstates to the destruction
operator ($ \hat{a}|\beta\rangle$ $\!=$ $\!\beta|\beta\rangle$), it 
is obvious that subtracting photons from a coherent state yields 
again a coherent state. 
Photon-added coherent states are highly 
nonclassical states. From
\begin{eqnarray}
\label{1.14}
(\hat{a}^\dagger)^{n}|\beta\rangle
& = &(\hat{a}^\dagger)^{n}\hat{D}(\beta)|0\rangle
\nonumber \\ 
& = &\hat{D}(\beta)\hat{D}^\dagger(\beta)
(\hat{a}^\dagger)^{n}\hat{D}(\beta)|0\rangle,
\end{eqnarray}
we find that
\begin{eqnarray}
\label{1.14c}
\nonumber
(\hat{a}^\dagger)^{n}|\beta\rangle &=&
\hat{D}(\beta)(\hat{a}^\dagger+\beta^\ast)^{n}|0\rangle\\ 
&=& \sum_{k=0}^{n}{n\choose k}\sqrt{k!}\,
(\beta^{\ast})^{n-k}\hat{D}(\beta)|k\rangle.
\end{eqnarray}
Equation (\ref{1.14c}) reveals that
photon-added coherent states are finite superpositions of  
displaced Fock states $\hat{D}(\beta)|n\rangle$ \cite{Agarwal2}
(for the properties of displaced Fock states, see 
\cite{Boiteux,Roy,Wuensche0}. From equations (\ref{2.3}) and (\ref{2.4b}) 
we know that PSJP and PAJP coherent states can be given by finite 
superpositions of photon-added coherent states. Expressing the 
latter, according to equations (\ref{1.14}) and (\ref{1.14c}), in 
terms of displaced Fock states, we see that PSJP and PAJP states
are also finite superpositions of displaced Fock states. 

Similarly, displaced Fock states can be given by finite
superpositions of photon-added coherent states. 
To show this, we write 
\begin{eqnarray}
\label{1.14b}
\nonumber
\hat{D}(\beta)|n\rangle
&=&\hat{D}(\beta)\frac{1} {\sqrt{n!}} \,(\hat{a}^\dagger)^{n}|0\rangle\\
&=&\frac{1}{\sqrt{n!}}\hat{D}(\beta)(\hat{a}^\dagger)^{n}
\hat{D}^\dagger(\beta)\hat{D}(\beta)|0\rangle,
\end{eqnarray}
and hence
\begin{eqnarray}
\label{1.14d}
\nonumber
\hat{D}(\beta)|n\rangle
&=& \frac{1}{\sqrt{n!}}(\hat{a}^\dagger -\beta^\ast)^{n}
|\beta\rangle\\
&=&
\frac{1}{\sqrt{n!}}\sum_{l=0}^{n}{n\choose l}(-\beta^{\ast})^{n-l}
(\hat{a}^\dagger)^l |\beta\rangle.
\end{eqnarray}
Note that from equation (\ref{1.14d}) and equation (\ref{2.4b})
for $|T|^2$ $\!=$ $\!|R|^2$ $\!=$ $\!0.5$ and $\beta$ $\!\approx$
$\!1$ it follows that $|\Psi_{11}\rangle$ $\!\sim$ 
$\!\hat{D}(\beta)|n\rangle|_{n=1}$, i.e., 
the conditional measurement schemes realizes a coherently
displaced single-photon Fock state.


\subsection{Near-photon-number eigenstates}
\label{subsec3.3}

Near-photon-number eigenstates are an example 
of minimum uncertainty states that are defined by the eigenstates of the 
operator $\hat{Y}$ $\!=$ $\!\hat{n}$ $\!-$ $\!i|\beta|\hat{x}(\varphi)$, 
which is associated with the ``simultaneous'' measurement of photon 
number and quadrature components \cite{Yuen2}. The states are
also called crescent states and have been studied in a number of papers
\cite{Yuen2,Hradil,Luks1,Luks2,Marian1,Agarwal1}. In particular, 
they can be expressed in the form of
\begin{equation}
\label{14a}
|\psi\rangle\sim(\hat{a}^\dagger+\beta^\ast)^{n}|\beta\rangle,
\end{equation}
or alternatively, in terms of nonunitarily shifted Fock
states, and it was shown that they can be generated by state
reduction via photon-number conditional measurement in 
nondegenerate parametric down conversion \cite{Agarwal1}.

\begin{figure}[h]
\centering\epsfig{figure=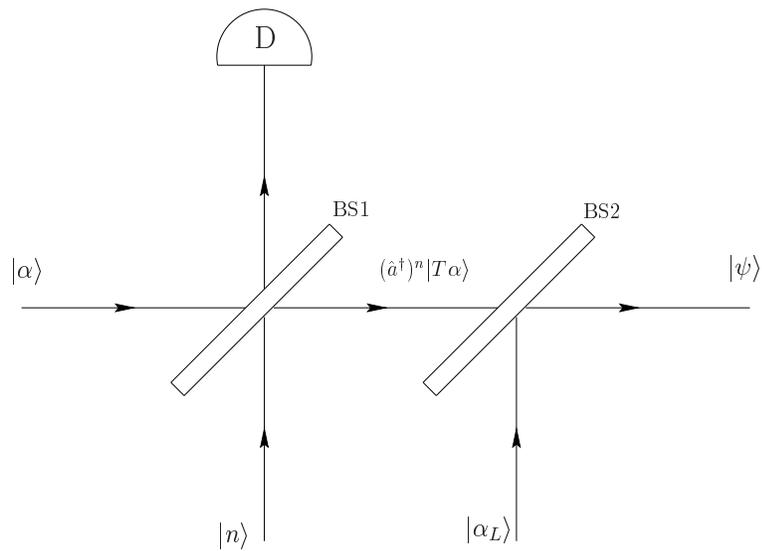,width=0.7\linewidth}
\caption{Experimental setup for preparing near-photon-number states
(crescent states). A mode prepared in a coherent state $|\alpha\rangle$ 
is first mixed (beam splitter BS1) with a mode prepared in a Fock state 
$|n\rangle$ and a zero-photon output measurement is performed (detector D).
The output mode prepared in a photon-added coherent state 
$(\hat{a}^\dagger)^n|2\beta\rangle$ ($2\beta$ $\!=$ $\!T\alpha$) is 
then mixed (beam splitter BS2 with transmittance $\tilde{T}$ close to unity) 
with a mode prepared in a strong coherent state $|\alpha_L\rangle$ such 
that $\beta$ $\!=$ $\!-(\tilde{R}/\tilde{T})\alpha_L$. }
\label{Fig6}
\end{figure}

   From equation (\ref{14a}) it is easily seen that
\begin{equation}
\label{14e}
|\psi\rangle
\sim \hat{D}^\dagger(\beta)
(\hat{a}^\dagger)^{n}\hat{D}(\beta)|\beta\rangle
=\hat{D}(-\beta)(\hat{a}^\dagger)^{n} |2\beta\rangle.
\label{1.14a}
\end{equation}
Equation (\ref{1.14a}) reveals that near-photon-number eigen\-sta\-tes 
are coherently displaced photon-added coherent states, which offers 
the possibility of producing them by conditional measurement on a beam 
splitter and subsequent coherent displacement. 
As it is depicted in Figure \ref{Fig6}, a mode prepared in a 
coherent state $|\alpha\rangle$ is first mixed with a mode
prepared in a Fock state $|n\rangle$, and a 
zero-photon measurement is performed in one output channel of the
beam splitter to prepare the
mode in the other output channel in a photon-added coherent state 
$(\hat{a}^\dagger)^{n} |2\beta\rangle$ ($2\beta$ $\!=$ $\!T\alpha$). 
To realize the coherent displacement $\hat{D}(-\beta)$,
this mode and a strong local-oscillator mode prepared in state
$|\alpha_{L}\rangle$ are then superimposed by another, unbalanced beam
splitter of high transmittance $\tilde T$ and low reflectance $\tilde R$
such that $\beta$ $\!=$ $\!-(\tilde R/\tilde T)\alpha_{L}$.


\subsection{Squeezed Fock states}
\label{subsec3.4}

Next let us consider a photon-added squeezed vacuum state,
\begin{eqnarray}
\label{1.15}
(\hat{a}^\dagger)^{n} \hat{S}(\xi)|0\rangle
& = &\hat{S}(\xi)\hat{S}^\dagger(\xi)
(\hat{a}^\dagger)^{n}\hat{S}(\xi)|0\rangle
\nonumber \\ 
& = & (1-|\kappa|^2)^{-n/2}
\hat{S}(\xi)(\hat{a}^\dagger+\kappa^\ast\hat{a})^{n}|0\rangle
\quad
\end{eqnarray}
[$\hat{S}(\xi)$ $\!=$ 
$\!\exp\{-\frac{1}{2} [\xi(\hat{a}^{\dagger})^{2}$
$\!-$ $\!\xi^{\ast}\hat{a}^{2}]\}$, 
$\kappa$ $\!\equiv$ $\!\kappa(\xi)$ 
$\!=$ $\!-e^{i\varphi_{\xi}}\tanh|\xi|$,
$\xi$ $\!=$ $\!|\xi|e^{i\varphi_{\xi}}$].
Using standard ordering techniques for boson operators 
\cite{Vogel2}, we derive
\begin{eqnarray}
\label{1.16}
\lefteqn{
\hspace*{-3ex}
(\hat{a}^\dagger+\epsilon\hat{a})^{n}=\sum_{l=0}^{n} {n\choose l} 
(\epsilon)^{n-l}
}
\nonumber \\ && \hspace{3ex}\times\, 
\sum_{k=0}^{[\frac{l}{2}]}\frac{(2\epsilon)^k}{\sqrt{\pi}}{l\choose 2k}
\Gamma\!\left(k+{\textstyle\frac{1}{2}}\right)
(\hat{a}^\dagger)^{l-2k}\hat{a}^{n-l},
\label{normal}
\end{eqnarray}
which enables us to rewrite equation (\ref{1.15}) as
\begin{eqnarray}
\label{1.17} 
\lefteqn{
(\hat{a}^\dagger)^{n} \hat{S}(\xi)|0\rangle =
(1-|\kappa|^2)^{-n/2}
}
\nonumber \\ && \hspace{10ex}\times\,
\sum_{k=0}^{[\frac{n}{2}]} \frac{n!(\kappa^\ast)^k}
{2^kk!\sqrt{(n\!-\!2k)!}}\,\hat{S}(\xi)|n-2k\rangle.
\end{eqnarray}
Analogously, for the photon-subtracted squeezed vacuum states we 
find that
\begin{eqnarray}
\label{1.18}
\hat{a}^{m}\hat{S}(\xi)|0\rangle
& = & \hat{S}(\xi)\hat{S}^\dagger(\xi)
\hat{a}^{m}\hat{S}(\xi)|0\rangle
\nonumber \\ & = & 
\kappa(1-|\kappa|^2)^{-m/2}
\hat{S}(\xi)(\hat{a}^\dagger+\kappa^{-1}\hat{a})^{m}|0\rangle\,,
\qquad
\end{eqnarray}
and hence
\begin{eqnarray}
\label{1.19}
\lefteqn{
\hat{a}^{m}\hat{S}(\xi)|0\rangle =
\kappa(1-|\kappa|^2)^{-m/2}
}
\nonumber \\ && \hspace{10ex}\times\,
\sum_{k=0}^{[\frac{m}{2}]}\frac{m!\left(1/\kappa\right)^k}
{2^kk!\sqrt{(m-2k)!}}\,\hat{S}(\xi)|m-2k\rangle.
\quad
\end{eqnarray}
Equations (\ref{1.17}) and (\ref{1.19}) show that photon-added and 
phot\-on-subtracted squeezed vacuum states can be given by  
finite superpositions of squeezed Fock states $\hat{S}(\xi)|k\rangle$,
and it is worth noting that the two classes of states realize 
two classes of Schr\"{o}dinger-cat states \cite{Dakna1,Dakna3}. 
The extension of equations (\ref{1.17}) and (\ref{1.19}) to
photon-added squeezed coherent states 
$(\hat{a}^\dagger)^{n}\hat{D}(\beta)\hat{S}(\xi)|0\rangle$
and photon-subtracted squeezed coherent states
$\hat{a}^{m}\hat{D}(\beta)\hat{S}(\xi)|0\rangle$ 
is straight\-for\-ward. These two classes of states can be
given by finite superpositions of displaced squeezed Fock states
$\hat{D}(\beta)\hat{S}(\xi)|k\rangle$
(for the properties of displaced squeezed Fock states, see
\cite{Satya,Nieto1}).

Displaced squeezed Fock states $\hat{D}(\beta)\hat{S}(\xi)|n\rangle$
can be rewritten as
\begin{eqnarray}
\label{1.20b}
\lefteqn{
\hat{D}(\beta)\hat{S}(\xi)|n\rangle=
(n!)^{-1/2} \hat{D}(\beta)\hat{S}(\xi) (\hat{a}^\dagger)^{n}|0\rangle
}
\nonumber \\ && \hspace{2ex}=\,
(n!)^{-1/2} \hat{D}(\beta)
\hat{S}(\xi)(\hat{a}^\dagger)^{n}\hat{S}^\dagger(\xi)\hat{S}(\xi)|0\rangle
\nonumber \\ && \hspace{2ex}=\,
[(1-|\kappa|^2)^{n}n!]^{-1/2}\hat{D}(\beta)
(\hat{a}^\dagger-\kappa^\ast\hat{a})^{n}\hat{S}(\xi)|0\rangle.
\quad
\end{eqnarray}
   From equations (\ref{1.20b}) and (\ref{1.16}) it is seen that
displaced squeezed Fock states cannot be given by finite 
superpositions of the photon-added squeezed states in general.
Note that the displaced Fock states $\hat{D}(\beta)|k\rangle$ 
are finite superpositions of photon-added coherent states 
[see equation (\ref{1.14d})].


\subsection{Squeezed-state excitations}
\label{subsec3.5}

\begin{figure*}[tbh]
\centering\epsfig{figure=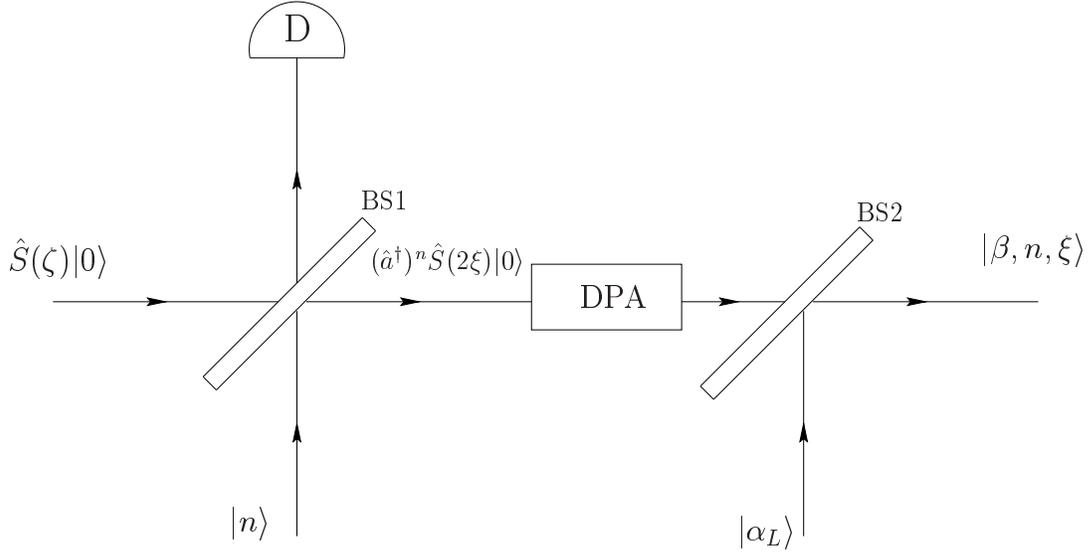,width=\linewidth}
\caption{Experimental setup for preparing squeezed-state
excitations. A mode prepared in a squeezed vacuum state 
$\hat{S}(\zeta)|0\rangle$ is first mixed (beam splitter BS1) with a 
mode prepared in a Fock state $|n\rangle$ and a zero-photon output 
measurement is performed (detector D), so that the output mode 
is prepared in a photon-added squeezed vacuum state $(\hat{a}^\dagger)^n
\hat{S}(2\xi|0\rangle$ [$\kappa(2\xi)$ $\!=$ $\!T^2\kappa(\zeta)$], 
which is then squeezed, e.g., by a degenerate parametric 
amplifier (DPA) to realize the squeezing parameter $-\xi$. Finally the
so squeezed output mode is mixed (beam splitter BS2 with transmittance 
$\tilde{T}$ close to unity) with a mode prepared in a strong coherent state
$|\alpha_L\rangle$ such that $\beta$ $\!=$ 
$\!(\tilde{R}/\tilde{T})\alpha_L$. }
\label{Fig7}
\end{figure*}

Squeezed-state excitations
\begin{equation}
\label{1.20}
|\beta,n;\xi\rangle
\sim \hat{D}(\beta) (\hat{a}^\dagger+\kappa^\ast\hat{a})^{n} 
\hat{S}(\xi)|0\rangle
\end{equation}
were introduced for diagonalizing the complete Gaussian class
of phase-space functions \cite{Wuensche1,Dodonov2}. Note that
although equation (\ref{1.20}) formally resembles
equation (\ref{1.20b}), squeezed-state excitations 
are quite different from displaced squee\-zed Fock states 
in general. It can be easily seen that
\begin{eqnarray}
\label{1.20c}
|\beta,n;\xi\rangle
& \sim & \hat{D}(\beta)\hat{S}^{\dagger}(\xi)\hat{S}(\xi)
(\hat{a}^\dagger+\kappa^\ast\hat{a})^{n}\hat{S}^{\dagger}(\xi)
\hat{S}(2\xi)|0\rangle
\nonumber \\
& \sim & \hat{D}(\beta)\hat{S}^{\dagger}(\xi)
(\hat{a}^\dagger)^{n}
\hat{S}(2\xi)|0\rangle
\end{eqnarray}
[cf. equation (\ref{1.15})].
We see that squeezed-state excitations are nothing but squeezed
and subsequently displaced pho\-ton-ad\-ded squeezed vacuum states,
which implies the sche\-me in Figure \ref{Fig7} for producing them.
Using a beam splitter, a mode prepared in a squeezed vacuum 
$\hat{S}(\zeta)|0\rangle$ is first mixed with a mode prepared in 
a Fock state $|n\rangle$ and a zero-photon measurement is performed 
in one output channel of the beam splitter to prepare the
mode in the other output channel in a photon-added squeezed
vacuum state $(\hat{a}^\dagger)^{n}\hat{S}(2\xi)|0\rangle$ 
[$\kappa(2\xi)$ $\!=$ $\!T^{2}\kappa(\zeta)$].  
This state is then squeezed (with $-\xi$ as squeezing parameter), 
e.g., by degenerate parametric amplification
or by state reduction via quadrature-component conditional
measurement in nondegenerate parametric down conversion 
\cite{Ban4}. Superimposing the outgoing mode and a strong local 
oscillator by an unbalanced beam splitter, the coherent displacement
can be realized [$\beta$ $\!=$ $\!(\tilde R/\tilde T)\alpha_{L}$;
cf. Section \ref{subsec3.3}].


\section{Realistic experimental conditions}
\label{sec6}

Let us first address the problem of realistic photon
detection. Unfortunately, there are no highly efficient and 
precisely discriminating photodetectors available at present. To overcome 
this  difficulty, photon chopping \cite{Paul1} was suggested 
for measuring the photon-number statistics.  
Let us remember that in such a scheme the mode to be detected is fed 
into an input channel of an optical $2N$-port array of beam splitters, 
the other \mbox{$N$ $\!-$ $\!1$} input ports being unused.
Highly efficient avalanche photodiodes in the $N$ output channels 
are used in order to record the coincidence event statistics.
Since they only distinguish between photons being present or absent,
the probability of obtaining $k$ clicks when $m$ photon are
present is given by \cite{Paul1}
\begin{equation}
\label{4.1}
\tilde P_{N,\eta}(k|m) = \sum_{l} \tilde P_{N}(k|l) \, M_{l,m}(\eta) 
\end{equation} 
($\eta$ is the detection efficiency), where
\begin{equation}
\label{4.2}
\tilde P_{N}(k|m) = \frac{1}{N^{m}} {N \choose k}
\sum_{l=0}^{k}(-1)^{l} {k \choose l} (k - l)^{m}  
\end{equation} 
for $k$ $\!\le$ $\!m$, and $\tilde P_{N}(k|m)$ $\!=$ $\!0$ for 
$k$ $\!>$ $\!m$. The matrix 
\begin{equation}
  \label{4.4}
  {M}_{l,m}(\eta) = {m \choose l} \eta^{l} (1 - \eta )^{m-l}
\end{equation}
for $l$ $\!\le$ $\!m$, and ${M}_{l,m}(\eta)$ $\!=$ $\!0$ for 
$l$ $\!>$ $\!m$ represents the effect of nonperfect detection. 
Since detection of $k$ coincident events can result from various numbers
$m$ of photons, the conditional state is in general a statistical mixture.
Therefore in place of equation (\ref{1.09}) we now have
\begin{eqnarray}
\label{4.5}
\hat \varrho_{\rm out}(n,k)
= \sum_{m,\Phi} \tilde{p}_{\Phi} P_{N,\eta}(n,m|k) \,|\Psi_{n,m} \rangle\langle \Psi_{n,m} |,
\end{eqnarray}
where $|\Psi_{n,m}\rangle$ is given in equation (\ref{1.10}), and
$P_{N,\eta}(n,m|k)$ is the probability of $m$ photons being present
under the condition that $k$ coincidences are recorded. The conditional
probability $P_{N,\eta}(n,m|k)$ can be obtained using the Bayes rule as
\begin{eqnarray}
\label{4.6}
P_{N,\eta}(n,m|k) = \frac{1}{\tilde P_{N,\eta}(n,k)} \tilde P_{N,\eta}(k|m)
\ P(n,m).
\end{eqnarray}
Here $P(n,m)$ is the prior probability (\ref{1.07}) of
$m$ photons being present, and accordingly, $\tilde P_{N,\eta}(n,k)$
is the prior probability of recording $k$ coincident events,
\begin{eqnarray}
\label{4.7}
\tilde P_{N,\eta}(n,k) = \sum_{m} \tilde P_{N,\eta}(k|m)
\ P(n,m) .
\end{eqnarray}

\begin{figure}[tbh]
\centering\epsfig{figure=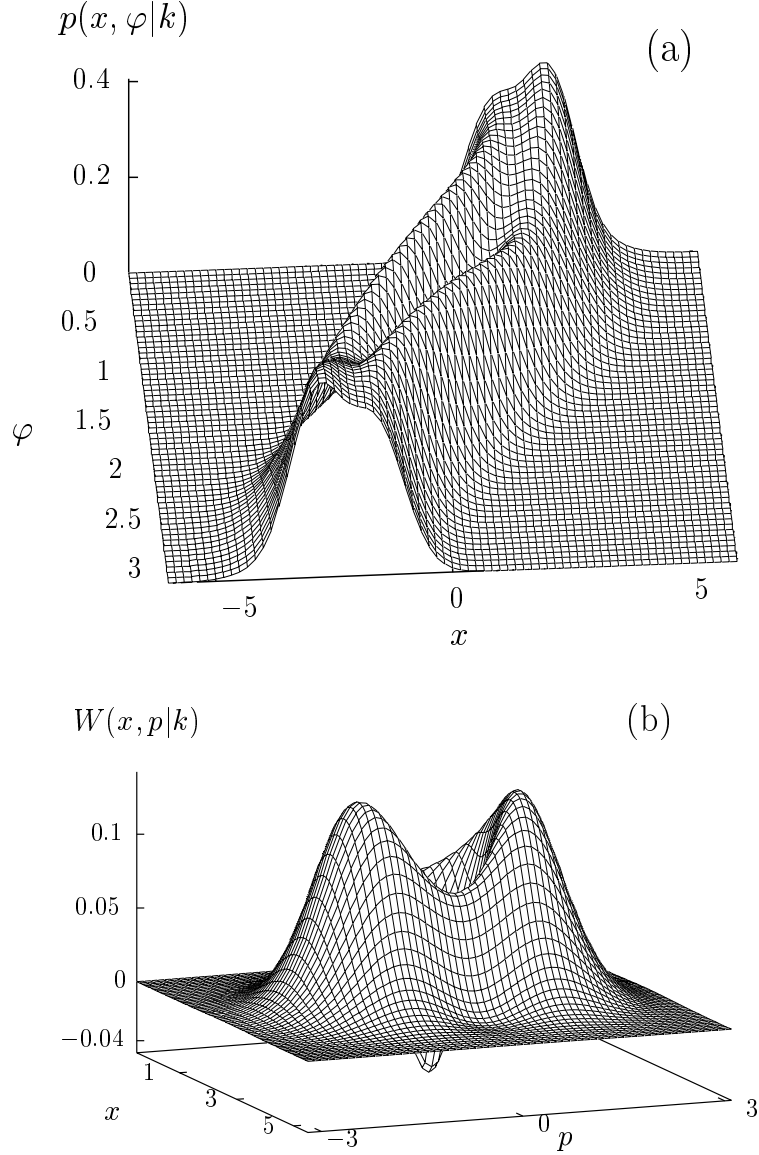,width=0.7\linewidth}
\caption{The quadrature-component distribution (a) and the Wigner 
function (b) of
a mixed conditional output state (\protect\ref{4.11}) for a 
coherent input state 
with $\beta'$ $\!=$ $\! 2.07$ 
($|\beta|$ $\!=$ $\! 2.3$, $|T|^2$ $\!=$ $\!0.81$). 
The parameters of the photon-chopping detection scheme are $k$ $\!=$ $\!4$,
$N$ $\!=$ $\!20$, $\eta$ $\!=$ $\!90$\% and those of the input 
Fock-state mixture (\protect\ref{4.9}) are $n_{0}$ $\!=$ $\!4$ 
and $p$ $\!=$ $\!0.95$. }
\label{Fig8}
\end{figure}

Second, preparation of the reference mode in a Fock state is 
a nontrivial problem (for a review, see \cite{Davidovich};
for single-photon Fock states, see also \cite{Hong1,Aspect1,Kwiat1};
for multiphoton Fock states, see also 
\cite{Kilin1,Law,Napoli,Steuernagel1,Paris1}).
In particular, a method for synthesizing multiphoton Fock states from 
sin\-gle-pho\-ton Fock states (produced, e.g., in parametric 
down conversion) has been proposed \cite{Steuernagel1,Paris1}. 
In the scheme, modes prepared in single-pho\-ton Fock states are fed into
the input ports of an array of beam splitters and 
detectors survey all but one output port so that the mode in the
free output port is prepared in the sought photon-number state. 
In practice however, it may be more realistic to consider 
statistical mixtures of photon-number states 
rather than pure Fock states. Let us return to equation (\ref{1.04}) 
and assume that
\begin{equation}
\hat \varrho_{\rm in} = \hat \varrho_{{\rm in}1}
\otimes \hat \varrho_{{\rm in}2},
\label{4.8}
\end{equation}
where 
\begin{equation}
\hat \varrho_{{\rm in}2} = \sum_{n} \tilde p_{n} \,
|n\rangle\langle n| \, .
\label{4.9}
\end{equation}
To be more specific, let us consider (as an example of a sub-Poissonian
distribution) a binomial probability 
distribution,
\begin{equation}
\tilde p_{n}
= {n_0\choose n} p^{n}(1-p)^{n_{0}-n} \quad {\rm if} 
\quad n\leq n_{0}
\label{4.10}
\end{equation}
and $\tilde p_{n}$ $\!=$ $\!0$ elsewhere ($0$ $\!<$ $\!p$ $\!<$ $\!1$).
Note that for \mbox{$p$ $\!\to$ $\!0$}, \mbox{$n_{0}$ $\!\to$ $\!\infty$}, 
and $pn_{0}$ finite the binomial distribution (\ref{4.10}) reduces to a 
Poisson distribution, with $\bar{n}$ $\!=$ $\!pn_{0}$ being the mean photon 
number. Using equations (\ref{4.8}) and (\ref{4.9}), from  equation (\ref{4.5}) we  
easily find that after recording $k$ coincident events  
the conditional mixed state now reads
\begin{eqnarray}
\lefteqn{
\hat{\varrho}_{{\rm out}1}(k)
= \sum_{n} \tilde p_{n}\,\hat{\varrho}_{{\rm out}1}(n,k)
}
\nonumber \\ && \hspace{8ex}= \, 
\sum_{n,m,\Phi} \tilde p_{n} \tilde p_{\Phi}\,P_{N,\eta}(n,m|k) 
\Psi_{n,m} \rangle \big\langle \Psi_ {n,m}|. 
\quad
\label{4.11}
\end{eqnarray}
Accordingly, the probability of detecting the state is 
the average of $P_{N,\eta}(n,k)$ given in equation (\ref{4.7}), i.e., 
\begin{eqnarray}
P_{N,\eta}(k) &=& \sum_{n} \tilde p_{n} P_{N,\eta}(n,k)\\&=&
\sum_{n,m}\tilde p_{n}\tilde{P}_{N,\eta}(k|m)P(n,m).
\label{4.12}
\end{eqnarray}
The quadrature-component distributions and the Wigner function
of a mixed conditional output state (\ref{4.11}) are plotted in figure 
\ref{Fig8} [$ P_{N,\eta}(k)$ $\!\approx$ $\!21.4\%$]. We see that 
the quantum interference is still preserved for realistic values 
of the number of photodiodes and efficiencies ($k$ $\!=$ $\!4$,
$N$ $\!=$ $\!20$,
$\eta$ $\!=$ $\!90$\%) and for a sub-Poissonian statistics of the 
input Fock-state mixture (\ref{4.9}) [$\bar{n}$ $\!=$ $\!3.8$, 
$\overline{(\Delta n)^{2}}$ $\!=$ $\!0.19$].  


\section{Summary and conclusions}
\label{sec7}
We have extended previous work on quantum-state preparation
via conditional output measurement on a beam splitter and
shown that when a mode prepared in a state $|\Phi\rangle$ is 
mixed with a mode prepared in an $n$-photon Fock state
and $m$ photons are detected in one of the output channels 
of the beam splitter, then the mode in the other output
channel is prepared in a photon-subtracted ($n$ $\!<$ $\!m$) or a
photon-added ($n$ $\!>$ $\!m$) Jacobi polynomial state which 
is obtained by applying an operator-valued Jacobi
polynomial to the state $|\Psi\rangle$ $\!\sim$
$T^{\hat{n}}|\Phi\rangle$. Since for typical classes of input states, 
such as thermal states, coherent states and squeezed states, 
the states $|\Phi\rangle$ and $|\Psi\rangle$ belong to the same 
class of states, the Jacobi polynomial states derived from
$|\Phi\rangle$ and $|\Psi\rangle$ also belong to the same class of states. 
Jacobi polynomial states are nonclassical states in general,
so that subtracting photons from or adding photons to them 
again yields nonclassical states in general. 

The analysis has shown that
PSJP and PAJP coherent states can be regarded as extremely non-Gaussian 
squeezed states. A characteristic feature of PSJP and PAJP squeezed 
vacuum states are the well pronounced quantum interferences associated 
with the quadrature-component noise reduction. Moreover,
these two classes of states represent Schr\"{o}dinger-cat-like states. 
Pho\-ton-ad\-ded Jacobi polynomial states are more nonclassical than
pho\-ton-sub\-trac\-ted Jacobi polynomial states in general. In particular
when $n$ $\!=$ $\!0$ \mbox{($m$ $\!>$ $\!0$)} or $m$ $\!=$ $\!0$ 
($n$ $\!>$ $\!0$), respectively, then the states reduce to 
ordinary photon-subtracted or photon-added states.
Whereas pho\-ton-ad\-ded coherent sta\-tes are non-Gaussian squeezed
states, subtracting photons from a coherent state obviously
leaves the state unchanged. 

The analysis has further shown that there are close relations
to other nonclassical states that have widely been studied.
Hence combining state preparation via con\-ditio\-nal 
output measurement on a beam splitter with other schemes
offers novel possibilities of nonclassical-state generation
and manipulation, such as the generation of 
near-photon-number eigenstates and squeezed-state excitations.
Since near-photon-number eigenstates are coherently
displaced photon-added coherent states, they can be 
generated by combining the scheme for photon adding with
a scheme for coherently displacing a state.
The latter can be realized by using a
second beam splitter whose transmittance is close to unity and
which mixes the mode prepared in a photon-added coherent state with 
a mode prepared in a strong coherent state. 
Similarly, a squeezed-state excitation can be prepared by
appropriately squeezing a photon-added squeezed vacuum state 
followed by a coherent displacement of the state. 

In order to demonstrate the feasibility of generating
PSJP and PAJP states, we have calculated the 
corresponding event probabilities. Further, we have also allowed 
for both nonprecise input Fock-state preparation and
nonprecise output photon counting. For this purpose we have considered 
sub-Poissonian mixtures of Fock states in place of pure Fock states
and assumed that photon-chopping is adopted for photon counting.
The results show that, apart from some smearing, typical properties 
of the states can still be observed even under realistic experimental 
conditions. 
  

\section*{Acknowledgement}
This work was supported by the Deutsche For\-schungs\-gemein\-schaft. 
We are grateful to E. Schmidt  and M.G.A. Paris for valuable 
discussions.


\begin{appendix}
\renewcommand{\thesection}{Appendix \Alph{section}:}
\section{Proof of equation (\protect\ref{jacobi})}
\label{app1}
\setcounter{equation}{0}
\renewcommand{\theequation}{\Alph{section}.\arabic{equation}}

To prove equation (\ref{jacobi}), let us consider the
operator function 
\begin{equation}
{\rm F}_{l}^{\mu}(\hat{n},|\nu|;T)
= T^{\hat{n}}\sum_{k=0}^{l}\frac{(-|R|^2)^{k}k!}{(k\!+\!|\nu|)!} 
{l\choose k}
{\hat{n}\!+\!\mu\!+\!k\choose k},
\label{jacobi1}
\end{equation}
which reads in the Fock basis as
\begin{eqnarray}
\label{A1}
\lefteqn{
{\rm F}_{l}^{\mu}(\hat{n},|\nu|;T)
=\sum_{n=0}^{\infty}
\frac{T^{n}}{(n\!+\!\mu)!}
}
\nonumber \\ && \hspace{2ex}\times\,
\sum_{k=0}^{l}
\frac{\Gamma(n\!+\!\mu\!+\!k\!+\!1)(-|R|^2)^{k}}
{\Gamma(k\!+\!|\nu|\!+\!1)} {l\choose k}|n\rangle\langle n| .
\end{eqnarray}
Introducing the integral representation of the gamma function 
$\Gamma(n$ $\!+$ $\!\mu$ $\!+$ $\!k$ $\!+$ $\!1)$, we have
\begin{eqnarray}
\label{A2}
\lefteqn{
\hspace*{-5ex}
{\rm F}_{l}^{\mu}(\hat{n},|\nu|;T)
=\sum_{n=0}^{\infty}\frac{T^{n}}{(n\!+\!\mu)!}
\int_{0}^{\infty} \! dt\,t^{n+\mu}e^{-t}
}
\nonumber \\ && \hspace{9ex}\times \, 
\sum_{k=0}^{l}\frac{(-|R|^2t)^{k}}{\Gamma(k\!+\!|\nu|\!+\!1)}{l\choose k}
|n\rangle\langle n|.
\end{eqnarray} 
We now use the sum rule (\ref{2.4a}) and the integral identity\cite{Prudnikov1}
\begin{eqnarray}
\label{A3}
\lefteqn{
\int_{0}^\infty dt\,t^{\alpha-1}e^{-t}{\rm L}_{n}^{\beta}(ct)
}
\nonumber \\ && \hspace{2ex}=\,
\Gamma(\alpha){\rm P}_{n}^{(\beta,\alpha-\beta-n-1)}(1-2c)
\quad({\rm Re} \ \alpha > 0)
\qquad
\end{eqnarray} 
and obtain ($|R|^2$ $\!=$ $\!1$ $\!-$ $\!|T|^2$)
\begin{eqnarray}
\label{A4}
\lefteqn{
{\rm F}_{l}^{\mu}(\hat{n},|\nu|;T)
}
\nonumber \\ && \hspace{1ex}=\,
\frac{l!}{(l\!+\!|\nu|)!}\sum_{n=0}^{\infty}
{\rm P}_{l}^{(|\nu|,n+\mu-|\nu|-l)}\!\left(2|T|^2\!-\!1\right)
T^{n}|n\rangle\langle n|
\nonumber \\ && \hspace{1ex}=\,
\frac{l!}{(l\!+\!|\nu|)!}
{\rm P}_{l}^{(|\nu|,\hat{n}+\mu-|\nu|-l)}\!\left(2|T|^2-1\right)
T^{\hat{n}}.
\end{eqnarray} 
Combining equations (\ref{1.21}), (\ref{1.20a}), (\ref{jacobi1})
and (\ref{A4}) eventually yields equation (\ref{jacobi}). 


\section{Photon statistics of PSJP and PAJP coherent states}
\label{app2}
\setcounter{equation}{0}
\renewcommand{\theequation}{\Alph{section}.\arabic{equation}}

  From equation (\ref{2.7}) the photon-number distribution 
$p_{n,m}(l)$ $\!=$ $\!|\langle\,l|\Psi_{n,m}\rangle|^2$
of PSJP and PAJP coherent states can be given by
\begin{eqnarray}
\lefteqn{
\hspace*{-2ex}
p_{n,m}(l)=\frac{e^{-|\beta'|^2}l!|\beta'|^{2l}}
{{\cal N}'_{n,m}|\beta'|^{2\nu}}
\Bigg |\sum_{k=\mu}^{n}(-|R|^2)^k {n\choose k}
}
\nonumber \\ && \hspace{16ex}\times \,
\sum_{j=\mu}^{k}{k\choose j}\frac{\theta(l-j)}
{(j\!-\!\nu)!(l\!-\!j)!}\Bigg |^2,
\label{2.7b}
\end{eqnarray}
where $\theta(n)$ $\!=$ $\!1$ for $n$ $\!\ge$ $\!0$ and 
$\theta(n)$ $\!=$ $\!0$ elsewhere. Further, from equations 
(\ref{2.2}) and (\ref{2.4}) it can be shown that the antinormally
ordered moments of the photon number can by given by
\begin{eqnarray}
\label{2.7f}
\lefteqn{
\hspace*{-6ex}
\langle \Psi_{n,m}|\hat{a}^p(\hat{a}^\dagger)^p|\Psi_{n,m}\rangle
=\frac{ |\beta'|^{-2\nu}}{{\cal N}'_{n,m}}
\sum_{k=\mu}^{n} \sum_{j=\mu}^{n}
(-|R|^2)^{k+j}
}
\nonumber \\ && \hspace{-7ex}\times   
{n\choose k}\!{n\choose j}
\!\sum_{l=\mu}^{k}\sum_{l'=\mu}^{j} \!{k \choose l}\!{j\choose l'}
\frac{\beta'^{l} (\beta'^{\ast})^{l'}
\chi_{l'+p,l+p}^{(1)}(\beta')}{(l-\nu))!(l'-\nu)!},
\end{eqnarray}
with $\chi_{l,k}^{(1)}(\alpha)$ being defined in equation (\ref{2.6}).
Equation (\ref{2.7f}) can then be used to derive closed solutions for the
normally ordered moments of the photon number. In particular, writing
\begin{eqnarray}
\label{2.7c}
\langle \hat{n}\rangle
=\langle \Psi_{n,m}|\hat{a}^\dagger\hat{a}|\Psi_{n,m}\rangle
=\langle \Psi_{n,m}|\hat{a}\hat{a}^\dagger|\Psi_{n,m}\rangle-1
\end{eqnarray}
and applying equation (\ref{2.7f}), the mean number of photons 
is calculated to be 
\begin{eqnarray}
\label{2.7d}
\lefteqn{
\hspace*{-2ex}
\langle \hat{n}\rangle=\frac{ |\beta'|^{-2\nu}}{{\cal N}'_{n,m}}
\sum_{k=\mu}^{n}
\sum_{j=\mu}^{n}(-|R|^2)^{k+j}
{n\choose k}{n\choose j}
}
\nonumber \\ && \hspace{-2ex}\times \,
\sum_{l=\mu}^{k}\sum_{l'=\mu}^{j}{k \choose l}{j\choose l'}
\frac{\beta'^{l}(\beta'^{\ast})^{l'}
\chi_{l'+1,l+1}^{(1)}(\beta')}{(l-\nu))!(l'-\nu)!}-1.
\end{eqnarray}
In a similar way, closed solutions can also be found for higher-order
moments. For example, in order to determine the Mandel factor 
$Q$ $\!=$ $\!(\langle \hat{n}^2\rangle$ $\!-$ 
$\!\langle \hat{n}\rangle^2)/\langle \hat{n}\rangle$ $\!-1$, 
knowledge of $\langle \hat{n}^2\rangle$ is required.
It can be obtained by introducing the antinormally ordered form 
\begin{eqnarray}
\label{2.7e}
\lefteqn{
\hspace*{-3ex}
\langle \hat{n}^2\rangle =
\langle \Psi_{n,m}|(\hat{a}^\dagger\hat{a})^2|\Psi_{n,m}\rangle
}
\nonumber \\ && \hspace{-2ex}= \, 
\langle \Psi_{n,m}|\hat{a}^2(\hat{a}^\dagger)^2|\Psi_{n,m}\rangle
-3\langle \Psi_{n,m}|\hat{a}\hat{a}^\dagger|\Psi_{n,m}\rangle + 1,
\quad
\end{eqnarray}   
and then applying equation (\ref{2.7f}).


\section{Derivation of equation (\protect\ref{2.18})}
\label{app3}\setcounter{equation}{0}
\renewcommand{\theequation}{\Alph{section}.\arabic{equation}}

According to equation (\ref{1.07}), the probability of producing  
PSJP and PAJP coherent states is given by 
\begin{eqnarray} 
\lefteqn{
\hspace*{-3ex}
P(n,m)=e^{-|\beta|^2}\frac{|R|^{-2\nu}n!}{|T|^{2m}m!}
}
\nonumber \\ && \hspace{-3ex}\times\,
\sum_{k=\mu}^{n}\sum_{j=\mu}^{n}(-|R|^2)^{k+j}
{m\choose k\!-\!\nu}{m\choose j\!-\!\nu}
\chi^{(2)}_{i,j}(\beta',\nu),
\quad
\label{2.17}
\end{eqnarray}
where
\begin{eqnarray}
\chi^{(2)}_{k,j}(\beta',\nu)
= \sum_{p=\delta}^{\infty}{p\!+\!k\choose k}
{p\!+\!j\choose j}\frac{|\beta'|^{2p}}{(p\!+\!\nu)!} \, ,
\label{2.17a}
\end{eqnarray}
which for $\nu$ $\!\geq$ $\!0$ (i.e., 
$\mu$ $\!=$ $\!\nu$ and $\delta$ $\!=$ $\!0$) can be rewritten as
\begin{eqnarray}
\lefteqn{
\hspace*{-3ex}
\chi^{(2)}_{k,j}(\beta',\nu)=
\frac{1}{k!\nu!}
}
\nonumber \\ && \hspace{-3ex}\times\,
\frac{\partial^k}{\partial(|\beta'|^2)^k}|\beta'|^{2k}
\left\{\frac{\nu !}{j!}
\sum_{n=0}^{\infty}\frac{\Gamma(n\!+\!j\!+\!1)}{\Gamma(n\!+\!\nu\!+\!1)}
\frac{|\beta'|^{2n}}{n!}\right\}.
\label{ap01}
\end{eqnarray}
The term in the curly brackets is nothing but the series expansion of
the confluent hypergeometric function. We therefore have
\begin{equation}
\chi^{(2)}_{k,j}(\beta',\nu)
=\frac{1}{k!\nu!}\frac{\partial^k}
{\partial(|\beta'|^2)^k}|\beta'|^{2k}\Phi(j+1,\nu+1,|\beta'|^2).
\label{ap02}
\end{equation} 
Using the relations \cite{Prudnikov1}
\begin{eqnarray}
\nonumber\Phi(a,b,z)=e^z\Phi(b-a,b,-z),\\[0.3cm]
\Phi(-n,b+1,z)=\frac{n!\Gamma(b+1)}{\Gamma(n\!+\!b\!+\!1)}{\rm L}^{b}_n(z),
\label{ap03}
\end{eqnarray}
we then obtain
\begin{eqnarray}
\lefteqn{
\hspace*{-4ex}
\chi^{(2)}_{k,j}(\beta',\nu)
= \frac{(j\!-\!\nu)!}{k!j!}\frac{\partial^k}
{\partial(|\beta'|^2)^k}|\beta'|^{2k}e^{|\beta'|^2}
{\rm L}^{\nu}_{j-\nu}(-|\beta'|^2)
}
\nonumber \\ && \hspace{-5ex}\!=\! 
\frac{(j\!-\!\nu)!}{j!}\sum_{l=0}^{k}{k\choose l}\frac{|\beta'|^{2l}}{l!}
\frac{\partial^l}{\partial(|\beta'|^2)^l}e^{|\beta'|^2}
{\rm L}^{\nu}_{j\!-\!\nu}(-|\beta'|^2).
\quad
\label{ap04}
\end{eqnarray}
The form (\ref{2.19}) of $\chi^{(2)}_{k,j}(\beta',\nu)$
used in equation (\ref{2.18}) follows by applying standard
formulas for derivatives of Laguerre polynomials \cite{Prudnikov1}.

For $\nu$ $\!<$ $\!0$ (i.e., $\mu$ $\!=$ $\!0$ and $\delta$ $\!=$ $\!-\nu$) 
equation (\ref{2.17a}) can be rewritten as
\begin{eqnarray}
\lefteqn{
\hspace*{-2ex}
\chi^{(2)}_{k,j}(\beta',\nu)
 \sum_{n=0}^{\infty}{n\!-\!\nu\!+\!k\choose k}
{n\!-\!\nu\!+\!j\choose j}\frac{|\beta'|^{2(n-\nu)}}{n!}
}
\nonumber \\ && \hspace{0ex}\!=\!  
\frac{1}{k!}\frac{\partial^k}{\partial(|\beta'|^2)^k}|\beta'|^{2(k-\nu)}
\sum_{n=0}^{\infty}{n\!-\!\nu\!+\!j\choose j}
\frac{|\beta'|^{2n}}{n!} \, .
\qquad
\label{ap05}
\end{eqnarray}
Using the relations \cite{Prudnikov1}
\begin{eqnarray}
\nonumber {\rm L}_{j}^{l-\nu}(0)
={n\!-\!\nu\!+\!j\choose j} , \\[0.3cm]
\sum_{n=0}^{\infty}{\rm L}_{j}^{n+b}(x)\frac{z^n}{n!}=
e^z{\rm L}_{j}^{b}(x\!-\!z),
\label{ap06}
\end{eqnarray}
we find that
\begin{eqnarray}
\lefteqn{
\hspace*{-5ex}
\chi^{(2)}_{k,j}(\beta',\nu)
= \frac{1}{k!}
\frac{\partial^k}{\partial(|\beta'|^2)^k}|\beta'|^{2(k-\nu)}e^{|\beta'|^2}
{\rm L}^{-\nu}_{j}(-|\beta'|^2)
}
\nonumber \\ && \hspace{-5ex}\!= \!
\sum_{l=0}^{k}{k\choose l}\frac{|\beta'|^{2l}}{l!}
\frac{\partial^l}{\partial(|\beta'|^2)^l}|\beta'|^{-2\nu}e^{|\beta'|^2}
{\rm L}^{-\nu}_{j}(-|\beta'|^2).
\label{ap07}
\end{eqnarray}
By again applying standard formulas for derivatives of the 
Laguerre polynomials \cite{Prudnikov1} 
equation (\ref{ap07}) takes the form used in equation (\ref{2.19}).
\end{appendix}


\newpage
\newpage
\newpage
\newpage
\newpage
\newpage
\newpage
\newpage
 
\end{document}